# Magnetooptics of layered two-dimensional semiconductors and heterostructures: progress and prospects

**Ashish Arora**

[†]Institute of Physics and Center for Nanotechnology, University of Münster, Wilhelm-Klemm-Strasse 10, 48149 Münster, Germany

Email: arora@uni-muenster.de

Fax: +492518333641

Beginning with the ‘conventional’ two-dimensional (2D) quantum wells (QWs) based on III-V and II-VI semiconductors in the 1970s, to the recent atomically-thin sheets of van der Waals materials such as 2D semiconducting transition metal dichalcogenides (TMDCs) and 2D magnets, the research in 2D materials is continuously evolving and providing new challenges. Magnetooptical spectroscopy has played a significant role in this area of research, both from fundamental physics and technological perspectives. A major challenge in 2D semiconductors such as TMDCs is to understand their spin-valley-resolved physics, and their implications in quantum computation and information research. Since the discovery of valley Zeeman effects, deep insights into the spin-valley physics of TMDCs and their heterostructures has emerged through magnetooptical spectroscopy. In this perspective, we highlight the role of magnetooptics in many milestones such as the discovery of interlayer excitons, phase control between coherently-excited valleys, determination of exciton-reduced masses, Bohr radii and binding energies, physics of the optically-bright and dark excitons, trions, other many-body species such as biexcitons and their phonon replicas in TMDC monolayers. The discussion accompanies open questions, challenges and future prospects in the field including comments on the magnetooptics of van der Waals heterostructures involving TMDCs and 2D magnets.

## 1. INTRODUCTION

The field of magnetooptics broadly deals with the interaction of electromagnetic waves with magnetism. Since many decades, it has played a central role in providing insights into the band-structure of solids[1–4]. On the one hand, magnetooptics is important for fundamental research[2], on the other hand, it is vital to many technological applications[1]. In this *perspective*, we highlight the crucial role played by magnetooptics in many new discoveries in the emerging area of layered two-dimensional (2D) materials involving van der Waals semiconductors and magnets. In the current section, we begin with a brief overview of magnetooptical effects from a historical perspective along with providing a glimpse into recent developments in the area of 2D materials. In section 2, we describe new spin-valley phenomena discovered in layered semiconductors under magnetic fields in the last few years, as outlined in Figure 1. For instance, valley Zeeman splitting and valley polarization under magnetic fields have contributed in various ways in identification and investigation of bright and dark excitons, trions, interlayer excitons, and four- and five-particle Coulomb-bound complexes in 2D. While a large part of the perspective covers Zeeman effects in 2D semiconductors (i.e. section 2.1 which is further divided into sections 2.1.1 till 2.1.6), other phenomena such as Landau quantization and diamagnetic shifts are discussed as well in appropriate details in sections 2.2 till 2.5, which have revealed important band-structure information such as exciton reduced masses, Bohr radii, and carrier magnetic moments. We also recapitulate the recent results involving magnetooptics of the emerging area of van der Waals heterostructures. Such heterostructures provide a rich platform for studying new physics in moiré potential landscape. In each subsection, we point out open questions and possible future directions of the field. In section 3, we conclude by highlighting the role and immense future potential of magnetooptics in the area of layered magnetic materials, and their heterostructures made with 2D semiconductors.

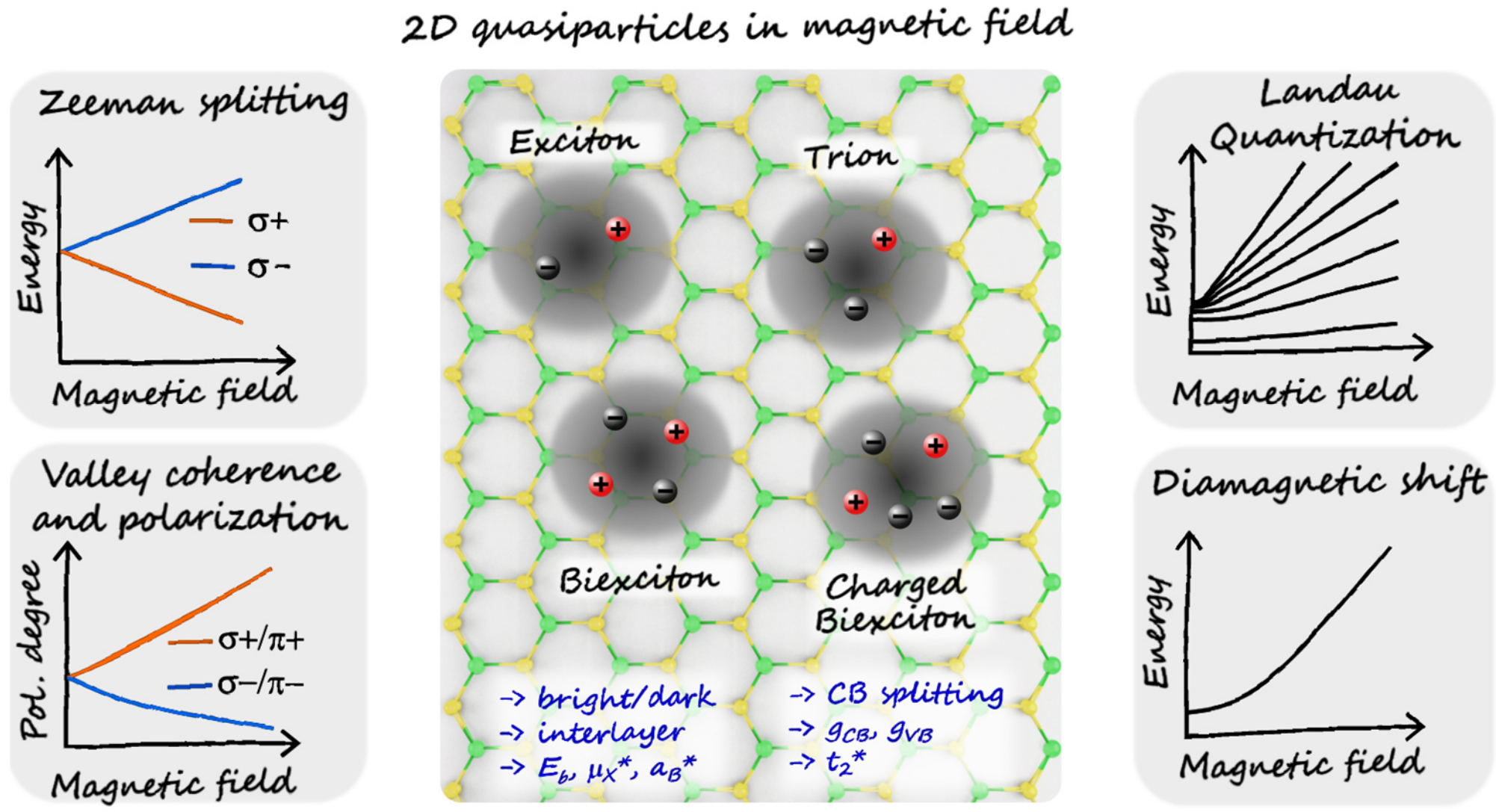

**Figure 1.** This perspective focuses on recent progress and future directions in the area of optical spectroscopy of 2-dimensional layered semiconductors under magnetic fields. Magnetooptical spectroscopy of Coulomb-bound quasiparticles such as excitons, trions, biexcitons and charged biexcitons is discussed in 2-dimensional systems. Titles of the gray panels describe the effects broadly covered in this perspective. Text in blue are some of the deduced quasiparticle parameters by studying these effects.

**A brief history of magnetooptics.** Magnetooptics was born after Michael Faraday's seminal works on the effects of a magnetic field on the polarization of light in the 1840s[5]. He noticed a rotation in the plane of polarization of light when it passed through a piece of lead-borosilicate glass placed under a magnetic field. The material under a magnetic field offers different refractive indices to the left- and right-circularly polarized components of the incident linear polarization, leading to the 'Faraday effect'. In 1876, John Kerr discovered a similar effect (magnetooptical Kerr effect or MOKE) while the light reflected off from a magnetized surface (an iron pole piece of a magnet)[6,7]. Augusto Righi found that in addition to the rotation of the state of polarization, the light becomes elliptically polarized in MOKE[8]. The resulting ellipticity is known as 'Kerr ellipticity' (or 'Faraday ellipticity' in transmission geometry). The effect was quantitatively measured by Pieter Zeeman[9]. In subsequent years, Zeeman discovered the much celebrated "Zeeman effect" by observing a splitting of Na lines in a magnetic field[10]. Since then, a variety of other magnetooptical effects have been discovered. Woldemar Voigt observed that a magnetic field applied transverse to the direction of propagation of light through Na vapor creates linear birefringence, known as 'Voigt effect'[11]. A similar effect was observed by Quirino Majorana in colloidal metal particles[12], and Aimé Cotton and Henri Mouton in paramagnetic liquids ('Cotton-Mouton effect')[13].

In the first half of the twentieth century, the Hanle effect was discovered which dealt with the reduction of degree of polarization of light emission from atoms under a transverse magnetic field, when they are excited using linearly polarized light[14,15]. Later on, this effect was used to determine spin relaxation times, nuclear fields and spin-orbit effects[16–20]. In the 1950s, the capability of MOKE/Faraday effect to observe magnetic domains[21], and to readout the magnetically-stored information (i.e. magnetooptic recording)[22,23] raised a lot of interest in magnetooptics. It was realized that an application of a magnetic field leads to Landau quantization, and reduces the dimensionality of the system by two.[24,25] An observation of cyclotron resonance of electrons and holes[26] and measurement of magnetoabsorption between quantized Landau levels[27,28] led to important experimental advancements to study quantum mechanical phenomena in solids. This was accompanied with many theoretical developments related to magnetoabsorption[29,30]. Roth et al. showed that the spin-orbit interaction can strongly modify the effective g factors of the electrons in a semiconductor leading to deviations from its free-space value of 2.[30] Interband Faraday rotation was also extensively studied theoretically, and experimentally[31–36]. In the 1970s, molecular beam epitaxy began producing high quality quantum wells (QWs) based on GaAs opening a new paradigm to investigate 2D physics using optical and transport methods[37]. By the 1980s, magnetooptical spectroscopy had become a powerful tool to study the electronic band structure of 2D quantum wells[2]. Rich low-dimensional physics of many-body quasiparticles such as excitons, trions and biexcitons was revealed[2,4,46–55,38,56–58,39–45].

The discovery of the giant magnetoresistance effect (GMR) around 1988 triggered the area of spintronics promising a new generation of energy-efficient and high-speed device applications. This brought the electron spin, a purely quantum-mechanical entity discovered in 1920s, on the forefront of modern and future technology and research. With the development of ultrafast-spectroscopy (sub picosecond) techniques around the same time[59–62], time-resolved magnetooptical spectroscopy opened up a new regime to study spin-relaxation processes and coherent spin manipulation in semiconductors and their QWs[63,64,73–82,65–72]. The area of research envisions spin based quantum-information processing, and quantum computation. The observation of the spin-Hall effect in thin films of GaAs and InGaAs established MOKE as an important tool for spintronics research[83].

Similar to the possibilities of spin manipulation in solids, a new area of research involving the manipulation of local minima in the momentum space i.e. valleys ("valleytronics") has recently emerged[84]. Although, ideas of valley manipulation date back to 1970s[85,86], atomically-thin sheets of layered semiconductors in the family of transition metal dichalcogenides (TMDCs) provide an ideal platform for studying the valley physics[87,88]. Monolayers of TMDCs of the type $MX_2$ ($M$=Mo, W; $X$=S, Se, Te) have energy-degenerate valleys at $K^{\pm}$ points of the Brillouin zone, which host the ground-state optical transitions[87–91]. The $K^{+}$ and $K^{-}$ valleys are connected by time-reversal symmetry, and couple to $\sigma^{+}$ and $\sigma^{-}$ circularly polarized light (Figure 2(b))[89–91]. This is because an absence of an inversion center combined with a strong spin-orbit interaction leads to a strong coupling ('locking') of the carrier spin to the valleys[89–91]. An external magnetic field applied perpendicular to the plane of the crystal breaks time-reversal symmetry, and lifts the valley degeneracy by shifting the two valleys in opposite directions as shown in

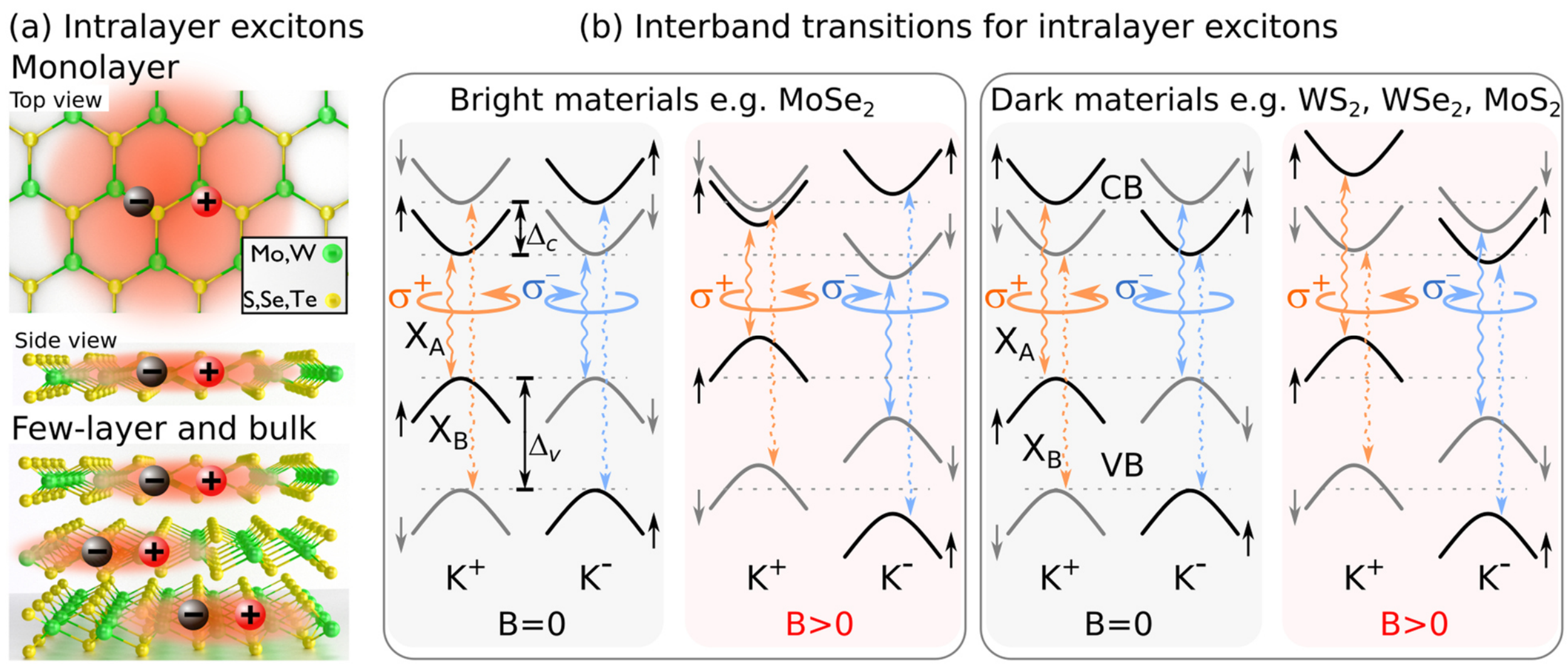

**Figure 2.** (a) Schematic representing intralayer excitons in a monolayer (top and side views), and a few-layer and bulk (side view) semiconducting transition metal dichalcogenide. (b) Interband A and B exciton 'bright' (optically allowed) $\sigma^+$ and $\sigma^-$ transitions for the intralayer excitons in monolayer and few-layer bright (left) and dark (right) materials are shown in the absence and presence of magnetic field. These transitions occur between the conduction and valence bands of the same spin. Horizontal dashed lines are the energies of the band extrema at $B = 0$. The energetic movement of the bands in the presence of a magnetic field is represented following the conventions in Ref. 146. The expected exciton g factor is about – 4 from this atomic picture (more details in text).

Figure 2(b) ("valley Zeeman splitting")[87,92]. This opens doors to control phase,[93–98] and population between the valleys[87,99–106]. More recently, intrinsic spin ordering (ferromagnetism and antiferromagnetism) has been discovered in new families of layered materials down to the atomically thin limit[107–110]. Some of the examples of these 2D magnetic materials (2DMMs) include thiophosphates of the form $MPX_3$ ($M$ = Mn, Fe, V, Zn, Co, Ni, Cd, Mg; $X$ = S, Se)[111,112], chromium trihalides of the form $CrX_3$ ($X$ = Cr, Br, I)[113,114], and others e.g. $Cr_2Ge_2Te_6$[115], $VSe_2$[116], $Fe_3GeTe_2$[117]. Heterostructures made from TMDCs and 2DMMs are promising to open new directions for valley manipulation using smaller magnetic fields through proximity effects[114].

Although, this *perspective* focuses on TMDCs and 2DMMs, research in the area of layered materials is continuously evolving. Nicolosi et al. explore the potential of liquid exfoliation techniques in many families of layered materials such as metal halides, oxides, III-VI layered semiconductors, layered $\alpha$ and $\gamma$ zirconium phosphates and phosphonates, layered silicates (clays), layered double hydroxides, and ternary transition metal carbides and nitrides[118]. Offering future directions in the field, computational techniques have sufficiently matured to predict hundreds of exfoliable and stable layered materials with desired characteristics to choose from[119,120].

## 2. 2D QUASIPARTICLES IN MAGNETIC FIELDS

Coulomb-bound electron-hole quasiparticles such as neutral and charged excitons, Fermi polarons, and biexcitons are important for the optical response of semiconductors[87,88,121–124]. Consider the case of an undoped semiconductor where neutral excitons dominate the optical spectra, especially when the binding energy of the exciton $E_b$ is comparable to or larger than the thermal energy scale given by $k_B T$ ($k_B$ is Boltzmann constant, $T$ being the temperature). In such a case, excitons appear as a strong resonance at an energy $E_{op}$ below the quasiparticle band gap $E_g$ by an amount equal to $E_b$ i.e. $E_{op} = E_g - E_b$. $E_{op}$ may also be called as the optical band gap of the material.

The exciton binding energy in a 2D semiconductor $E_b^{2D}$ is theoretically expected to be larger than the 3D case by a factor of 4 in Wannier approximation i.e. $E_b^{2D} = 4E_b^{3D}$.[125] Typically, the binding energies of excitons in conventional quantum wells ranges from a few meVs to a few 10s of meVs[46,47,122,123,125]. Mostly cryogenic temperatures are required to study the exciton ground and excited states in these systems. On the other hand, $E_b$ is much larger i.e. from few 10s of meVs to 100s of meVs in semiconducting TMDCs[87,88,126–131]. Therefore, TMDCs are an excellent platform to study these 2D hydrogen-like quasiparticles, as well as other many-body species such as charged excitons, Fermi polarons, biexcitons and charged biexcitons at elevated temperatures.

Some of the most important parameters which determine a Coulomb-bound composite's contribution to the optical response of a material are its binding energy, reduced mass $\mu^*$ and Bohr's radius. Magnetooptical methods are powerful tools to determine these parameters[2,4]. An externally applied magnetic field can affect the optical signatures in many ways: a) Zeeman splitting, b) Valley polarization and coherence effects, c) Landau quantization and oscillator strength enhancement, and d) Diamagnetic shift, e) Binding-energy enhancement. We discuss these effects one by one as follows.

**2.1 Zeeman splitting.** In this section, we discuss the Zeeman effects in 'bright' and 'dark' neutral and charged excitons, biexcitons, and interlayer excitons. Conduction and valence bands (CB and VB respectively) of a semiconductor undergo a Zeeman splitting under a magnetic field characterized by their effective g factors. The effective g factor of a carrier in a band is strongly influenced by the spin-orbit interaction-induced band mixing[68,132–135]. If the conduction and the valence bands which are involved in the exciton creation undergo Zeeman shifts of $\Delta E_{Zm}^{c}$ and $\Delta E_{Zm}^{v}$ under a magnetic field, then the Zeeman splitting of the exciton line can be approximated by[87]

$$\Delta E_{Zm}^{X} = 2(\Delta E_{Zm}^{c} - \Delta E_{Zm}^{v}) = (g_c - g_v)\mu_B B \cong g_X \mu_B B \quad (1)$$

where $g_c$, $g_v$ and $g_X$ are the effective CB, VB and exciton g factors, $\mu_B = 0.05788$ meV/T is the Bohr's magneton, and $B$ is the applied magnetic field[2]. The Zeeman-split exciton lines for bright excitons are circularly polarized and are detected separately by helicity-resolved photoluminescence (PL)[96,99,139–141,100–103,105,136–138], reflectance[87,104,142–146], transmittance[56,101,147] or photoconductivity spectroscopy[2,148]. Zeeman splitting in terms of circularly-polarized components is given as

$$\Delta E_{Zm}^{X} = E^{\sigma+} - E^{\sigma-} = g_X \mu_B B \quad (2)$$

Here $E^{\sigma+}$ and $E^{\sigma-}$ are the transition energies for the left and right circularly polarized light in the helicity-resolved spectra. Alternatively, polarization-modulation-spectroscopy measurements are performed such as Faraday rotation[35], MOKE spectroscopy[149,150], and magnetic circular dichroism (MCD)[151,152] to determine the Zeeman splitting. Other techniques such as quantum beats in time-resolved pump-probe spectroscopy[64], quantum beats in time-resolved PL in Voigt geometry[66,68–71,73–76,79], time-resolved Faraday spectroscopy in Faraday geometry[67,72,78], time-resolved Kerr spectroscopy[77,81], four-wave mixing[153], Spin-flip Raman scattering[154] and hole burning[65] have also been used to measure the carrier and exciton g factors in quantum wells.

We briefly revise the well-established Zeeman effects in QWs before moving on to the layered semiconductors. The g factors of excitons in QWs have been found to be strongly dependent upon the well thickness. For instance, the heavy-hole exciton g factor in GaAs QW is negative for small well widths (about −2 for a well width of 2 nm), while it changes sign for a well width of around 6 nm, and approaches a value of about +1 for large well widths (> 25 nm)[136,150]. The light-hole exciton g factor stays between 6 to 8 for this range of well widths[150]. The g factor of electrons has been studied extensively as a function of well width and it increases monotonically from the bulk GaAs value of −0.44 to about +0.4 with reducing well thickness to 3 nm[133,155,156]. Theoretical attempts have been made to calculate the g factors using $\mathbf{k} \cdot \mathbf{p}$ perturbation theory[133,134,157]. While a good agreement with the experimentally obtained electron and heavy-hole g factors was obtained[133,134], a satisfactory explanation of the trends for the light-hole g factors requires more work[158,159]. The monotonically increasing electron's and heavy-hole g factors in CdTe/$Cd_{1-x}(Mg_x)$Te QWs with decreasing well-width were also successfully modeled by $\mathbf{k} \cdot \mathbf{p}$ theory[160]. This method has also been used to explain

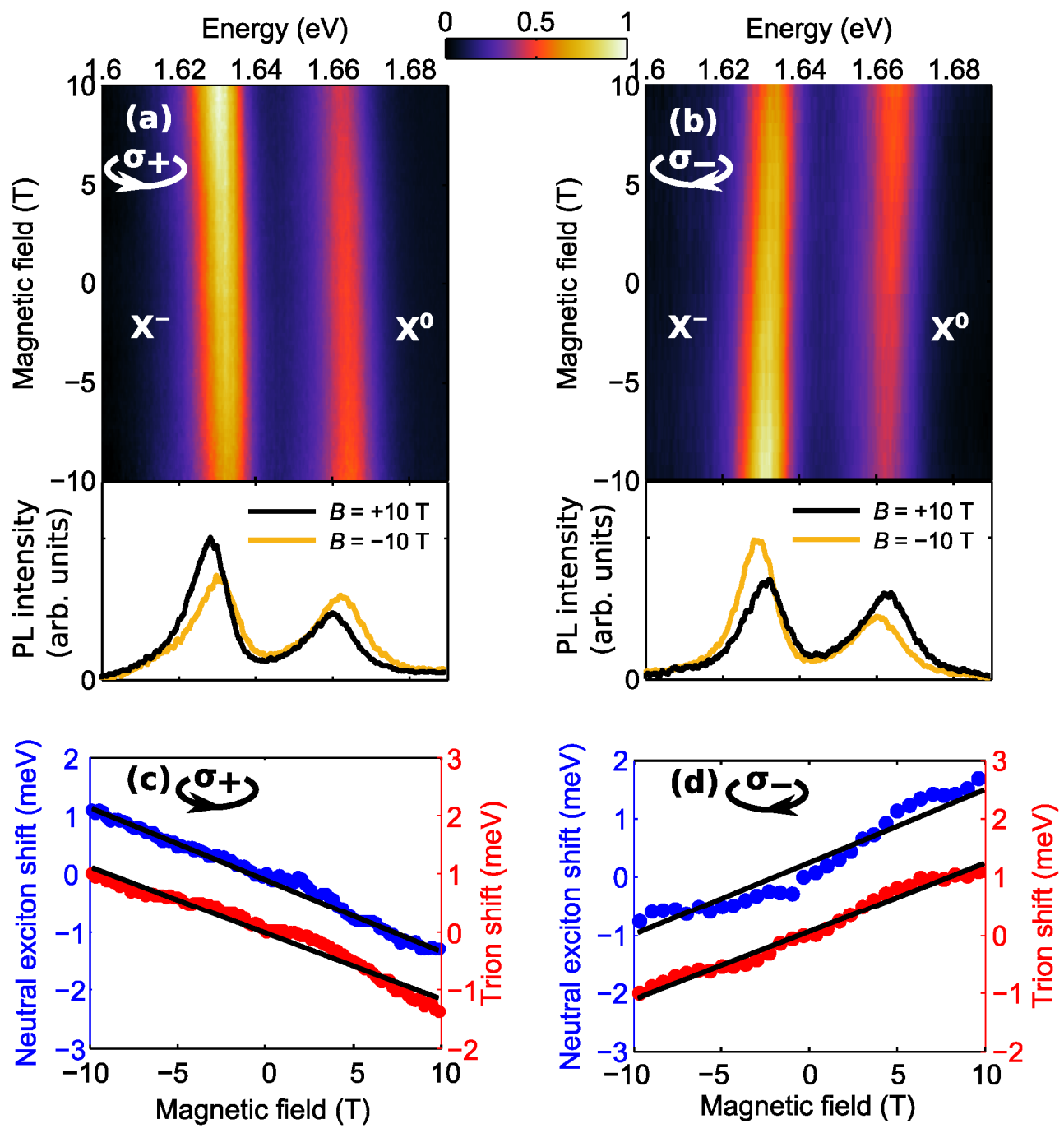

**Figure 3.** (a) and (b) False color maps of the helicity-resolved photoluminescence of monolayer $MoSe_2$ on Si/(300 nm)$SiO_2$ substrate under out-of-plane magnetic field from −10 T to 10 T. Panels below the color maps are PL spectra for both circular polarizations at $B = \pm 10$ T. Magnetic-field-induced variation of PL intensity signifies the valley polarization effect discussed in the text. (c) and (d) Neutral ($X^0$) and charged ($X^-$) excitons shift in energy with magnetic field corresponding to a g factor of about −4. See Ref. 99 for further details. Reproduced with permission from Li et al., Phys. Rev. Lett. 113, 266804 (2014). Copyright 2014 American Physical Society.

the increasing g factor of electrons in GaAs with rising temperature[135,161].

In the case of TMDCs, strong spin-orbit interaction leads to a splitting of the conduction ($\Delta_c$) and valence ($\Delta_v$) bands as shown in Figure 2(b). $\Delta_c$ is of the order of few meVs to a few 10s of meVs, while $\Delta_v$ ranges from about 150 meV to 500 meV[162–165]. Due to the specific sign of the conduction-band splitting, the ground-state exciton in $WS_2$, $WSe_2$, and $MoS_2$ is found to be optically "spin forbidden" or "dark"[87,162,172–181,164,182,165–171] (see Figure 2(b)). In $MoSe_2$ and $MoTe_2$, however, the ground-state exciton is an optically "bright" state due to a reversed conduction band splitting compared to the other three materials[165,182–185] (Figure 2(b)). The brightness or the darkness of the ground-state excitons have a strong influence on the temperature- and carrier-density-dependent optical and magnetooptical properties (absorption and PL) of TMDCs[166,167,183,186,187]. We first discuss the Zeeman effects of bright excitons, followed by a discussion on dark excitons.

***2.1.1. Neutral 'bright' excitons in TMDCs.*** For monolayer TMDCs, most of the works report the 'bright' A and B exciton g factors close to −4 using magnetoreflectance, transmittance and PL at liquid He temperatures, for out-of-plane applied magnetic fields[87,96,145–147,188,100–105,139,141]. As an example, Figure 3 shows the circular-polarization-resolved magneto PL of monolayer $MoSe_2$ as measured by Li et al.[99] Contrary to the case of III-V and II-VI QWs, a simplistic 'atomic picture' incidentally provides a qualitatively correct value of g factors in certain cases involving excitons in TMDCs. An example is the proposed explanation of the g factors of A and B excitons in monolayer TMDCs[100,146]. Here the g factors of the bands participating in the exciton formation are assumed to be equal to the sum of spin, orbital and valley moments. In this picture, the transition metal's $d_{x^2-y^2} \pm i\, d_{xy}$ hybridized orbitals at the top of the valence band at the $K^+$ and $K^-$ valleys in the Brillouin zone are mainly supposed to contribute to the exciton's g factor[99,100,103,104,139,146]. The magnetic moment associated with these orbitals is $\pm 2\mu_B$. Then the Zeeman splitting would be obtained as $-4\mu_B B$ yielding an effective excitonic g factor of $-4$. It has been argued that the contributions due to valley magnetic moment, which depends on the effective masses of the charge carriers forming an exciton, would lead to deviations from this value in a monolayer[99,100,103,139,142,146]. An attempt was made to find the spin, orbital and valley contributions to the exciton g factor separately, in monolayers of $WS_2$ and $MoSe_2$[146]. The valley term was found to significantly contribute to the energies of the conduction and valence band states with a g factor of $g_V = 1.5 \pm 0.5$ for the two materials[146].

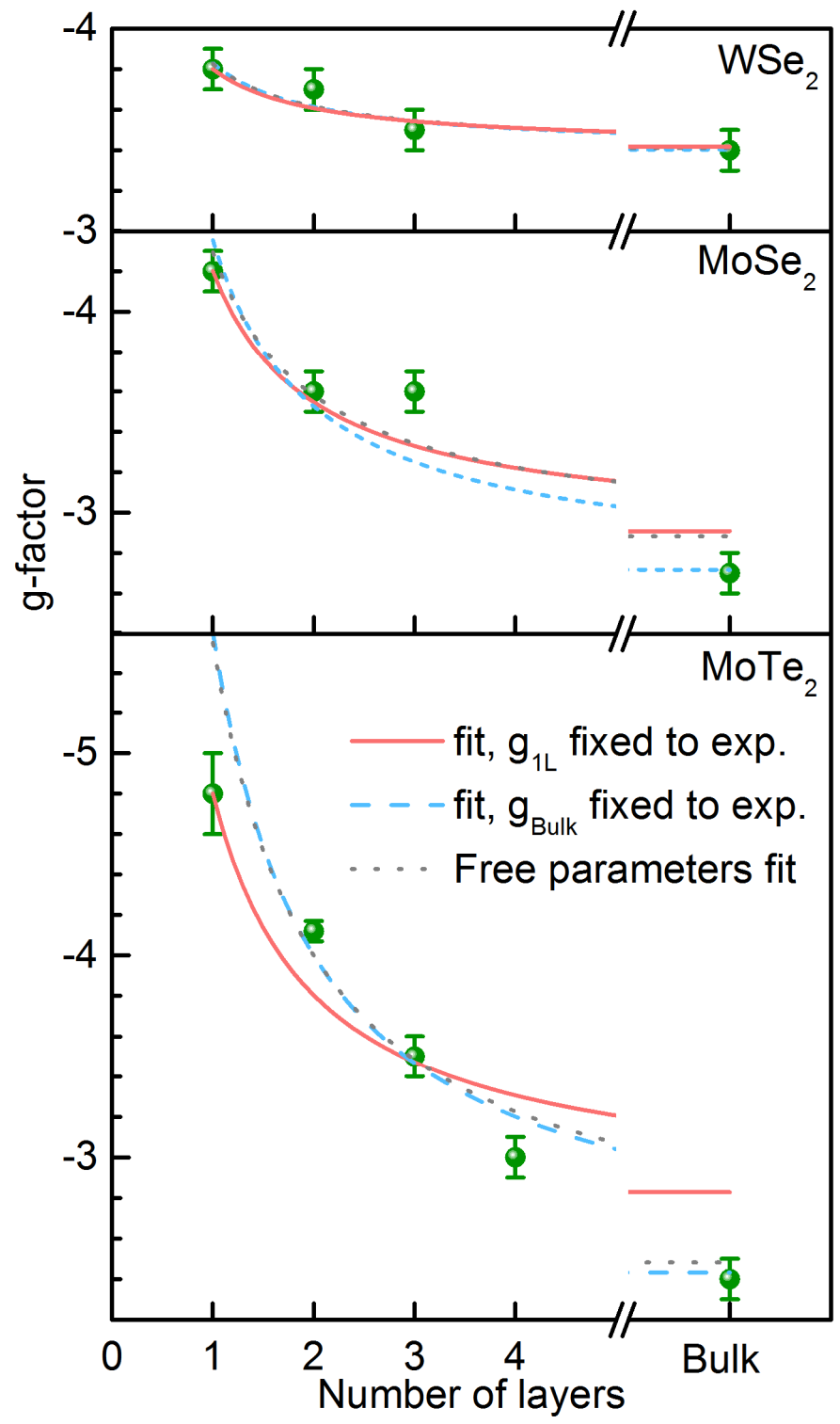

**Figure 4.** Effective g factors of the ground-state A exciton of a monolayer to the bulk limit in $WSe_2$, $MoSe_2$ and $MoTe_2$. Green spheres are the experimental data and lines are the fit using $\boldsymbol{k}\cdot\boldsymbol{p}$ perturbation theory[145]. Reproduced from Arora et al., 2D Mater. **6**, 015010 (2018). Copyright 2018 Institute of Physics Publishing. 

The fact that this atomic picture works for the A and B excitons could be a complete coincidence, since it does not explain the findings of the recent experiments[146,189] and *ab initio* theory[190–193] concerning g factors of conduction and valence bands in TMDCs. The authors of Ref. 146 determine the g factor for the lower conduction band equal to 1.08 and 1.84 for $WS_2$ and $MoSe_2$, respectively. A recent work reports the g factors of the bottom (top) conduction

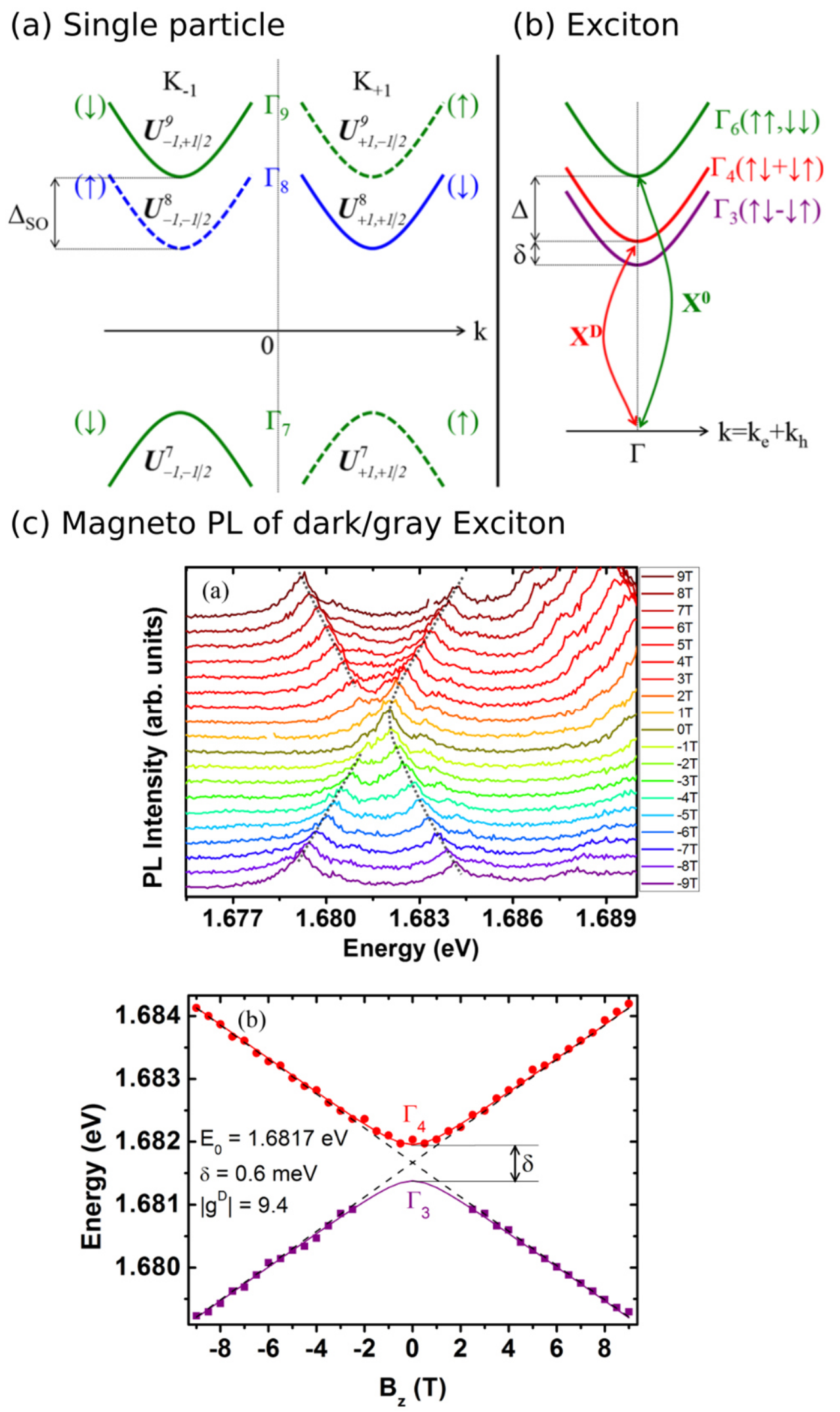

**Figure 5.** (a) Single-particle band structure of a monolayer $WSe_2$ for the $K^{\pm}$ valleys by choosing a complete set of electron eigenstates in the irreproducible representations of $D_{3h}$. (b) Band structure in the exciton picture at the Γ point of the exciton Brillouin zone. Exciton states belonging to the $\Gamma_6, \Gamma_4$ and $\Gamma_3$ representations are in-plane-dipole allowed ("bright"), out-of-plane-dipole allowed ("gray") and dipole forbidden ("dark"), respectively. The arrows represent the electric spin components. (c) Magneto-PL of the gray and dark excitons in the Faraday geometry. At 0 T, only gray exciton is visible. In the presence of a magnetic field, the gray and dark excitons may couple to each other and the dark exciton acquires a small oscillator strength making it possible to measure the exchange splitting $\delta = 0.6$ meV and a dark exciton g factor of 9.4. More details are found in Ref. 170. Reproduced with permission from Robert et al., Phys. Rev. B 96, 155423 (2017). Copyright 2017 American Physical Society.

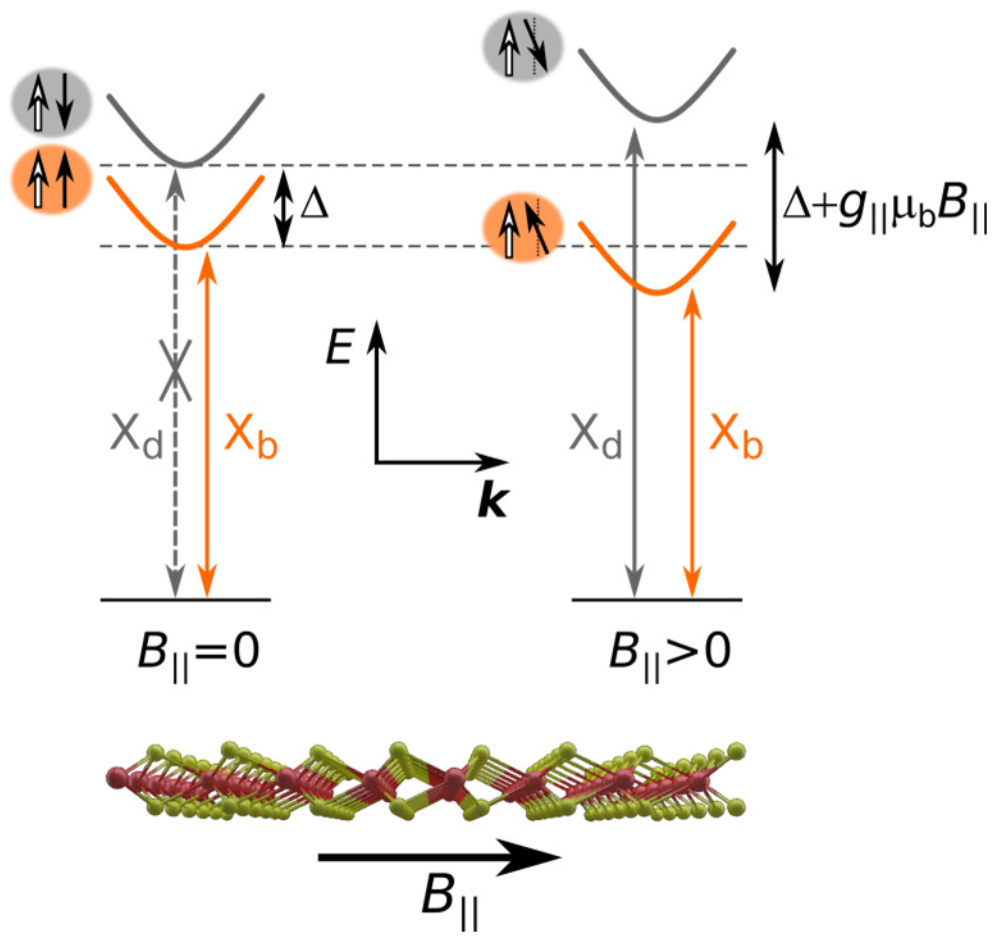

**Figure 6.** Schematic depicting dispersion of bright (orange color) and dark (grey color) excitons and an in-plane magnetic-field-induced mixing between them (Voigt geometry). Filled and empty arrows represent electrons and hole spins, respectively. In-plane magnetic field mixes the conduction band spins between bright and dark excitons. Therefore, the otherwise optically-disallowed dark excitons acquire some oscillator strength. The splitting between bright and dark excitons in this geometry is used in determining the electronic g factor of about 2.[182,185]

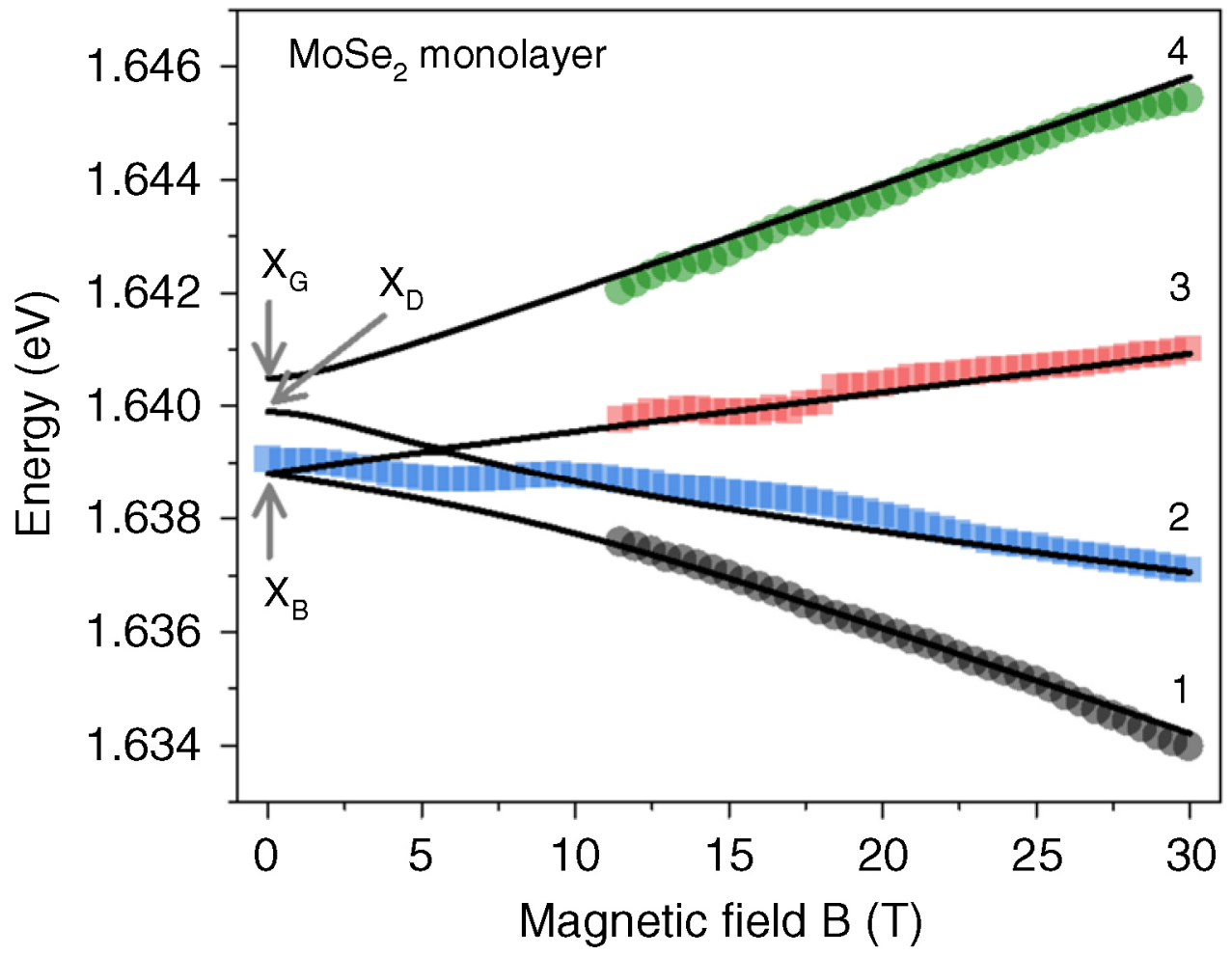

**Figure 7.** Energies of bright, dark and grey excitons deduced from magneto-PL spectra of an hBN encapsulated $MoSe_2$ monolayer under a tilted magnetic field (45° with respect to the sample plane) up to $B = 30$ T. The in-plane component of the magnetic field brightens the dark excitons, while the out-of-plane component separates them in energy (valley Zeeman splitting). Figure adapted from Robert et al., Nat. Commun. 11, 4037 (2020) (Creative Commons Attribution 4.0 International License)

bands as $0.86 \pm 0.1$ $(3.84 \pm 0.1)$ and for the top valence band as $6.1 \pm 0.1$[189] in an hBN-encapsulated $WSe_2$ monolayer. These values differ from the atomic picture involving simple additions of the spin, valley and orbital magnetic moments highlighting the requirement of a more complete theory. Recently, *ab initio* theory has matured to be able to calculate the g factors of the bands and excitons in TMDCs[190–193]. The values of the band g factors measured in [146] and [189] are in a good agreement with these *ab initio* calculations. Experiments on other TMDCs are required to be performed before their CB and VB g factors could be compared with the *ab initio* theory.

When the thickness of the TMDC is increased, it is found that the magnitude of the A exciton g factor reduces while staying negative (See Figure 4)[145]. This reduction is stronger for Mo-based TMDCs ($MoSe_2$ and $MoTe_2$ in Ref. 145) compared to $WSe_2$. The physics of thickness-dependent g factors of K-point excitons in TMDCs cannot be directly compared with their well-width dependence in quantum wells. It is because in TMDCs thicker than a monolayer, the neighboring layers are only weakly coupled to each other through van der Waals interactions[141,143,145,194–198]. Therefore, a multilayer TMDC acts more like a weakly coupled QW system for the states at the K point of the Brillouin zone, with every individual layer acting as a QW on its own. The K-point electron and hole wavefunctions are largely localized within a single layer of a multilayer TMDC[143,199]. Therefore, A and B excitons in a multilayer and a bulk TMDC largely possess two dimensionality[143,200]. When the thickness of the TMDC increases, $\mathbf{k} \cdot \mathbf{p}$ theory is able to model the reduction of the g factor of A excitons by incorporating interlayer-interaction-induced mixing of the conduction and valence states[145]. However, there is a scope for a more detailed theory for calculating the g factors such an based on *ab initio* methods. Furthermore, the thickness dependence of the B exciton g factors is not reported yet, and could provide more surprises[136,150].

One puzzle which is a topic of discussion in the community is that the g factor of A excitons in $MoS_2$ monolayer encapsulated with hBN has been found to deviate significantly from the value of $-4$,[182,201,202] where the values as low as $-1.7$ are obtained[201]. Authors of Ref. [202] find that the g factor in $MoS_2$ monolayers shows a systematic trend where the poor quality samples (larger exciton linewidths) show larger magnitudes of the g factor when compared to those exhibiting narrow linewidths. What needs to be checked is if this behavior is typical only to $MoS_2$, or to the other TMDCs as well. Another question one might ask is how the exciton and band g factors vary as one raises the temperature. The g factor of electrons in GaAs quantum wells is known to change from its low temperature value of $-0.44$ to about $-0.33$ at room temperature[68,135,161]. Interestingly, initial $\mathbf{k} \cdot \mathbf{p}$ calculations predicted the opposite trend[68]. However, later on it was realized that one needs to consider an integration over many Landau levels and over $k_z$ in the theory to explain the experimental results[135]. Such temperature-dependent measurements and theories have not been reported for TMDCs so far.

It is noteworthy that in the PL spectra of $WS_2$ and $WSe_2$ monolayers, many features lower in energy to the bright trions are observed, which were initially attributed to localized exciton states[203]. The origin of these peaks has been a topic of debate. Koperski et al. found that many of these PL lines have anomalous g factors ranging from -4 to -13[146]. This raised speculations that these lines could be related to the intravalley dark excitons with predicted g factors of −8, other shake-up processes involving a spin flip of excess electrons from low to upper conduction bands[146]. Recent magneto-PL experiments and theoretical developments have confirmed some of these speculations to be true. For instance, emission lines corresponding to the dark excitons[170,182,185,204–206], dark trions[176,178,181,205,207], phonon replicas of spin- and momentum-forbidden dark excitons[179,180], and phonon replicas of positive and negative dark trions[177,179,189,207,208] have been identified in monolayer $WSe_2$ magneto PL, on the bases of their observed g factors (ranging from -6 to -13). Another recent work also identifies these lines in hBN-encapsulated monolayer $WS_2$[209]. We discuss some of these features in the following.

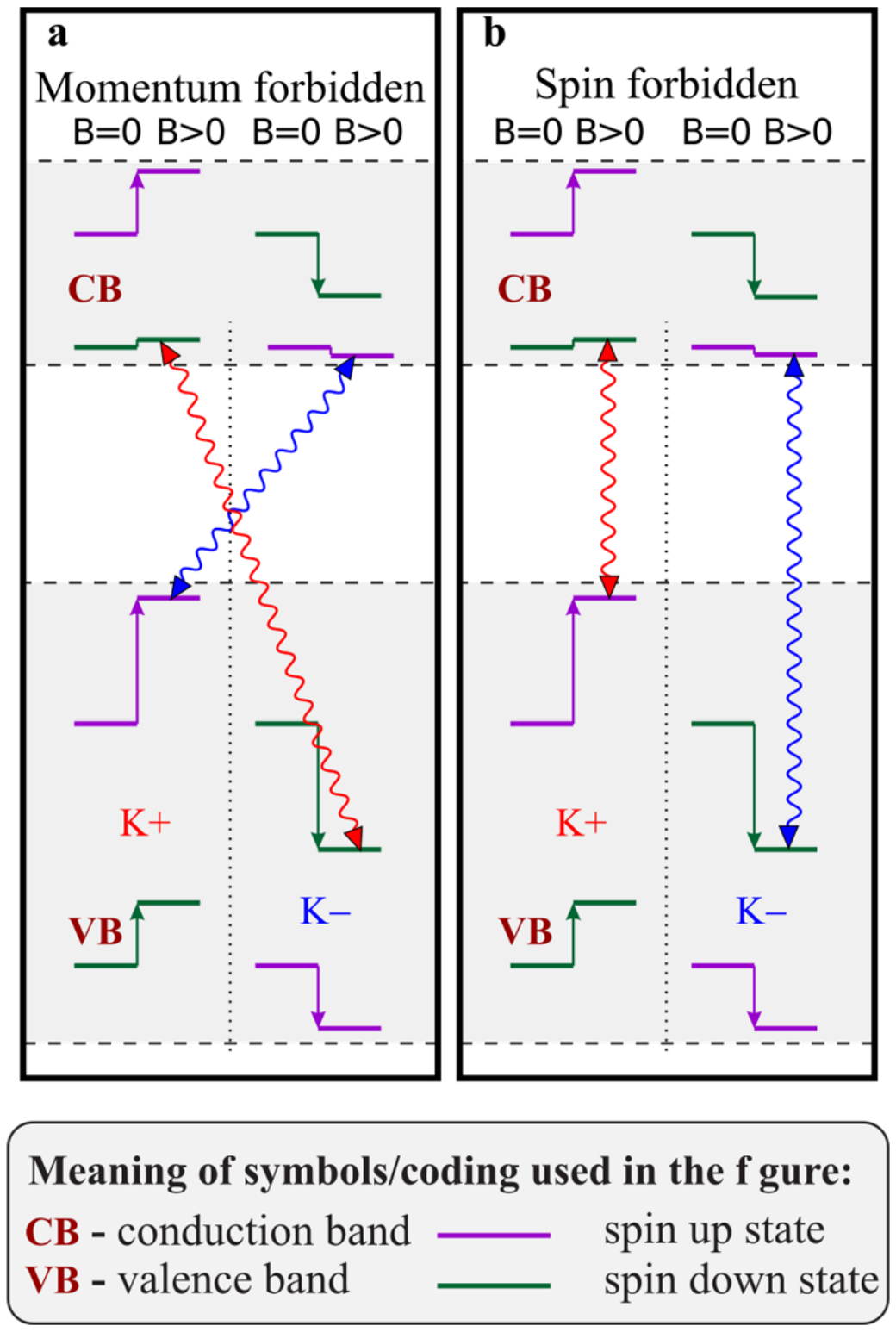

**Figure 8.** Examples of some (a) momentum-forbidden and (b) spin-forbidden transitions in dark TMDCs such as $WSe_2$ in the absence and presence of a magnetic field, as described in Ref. 146. These transitions are expected to exhibit much larger g factors than the bright exciton value of −4. For instance, the expected g factors of momentum and spin-forbidden transitions in this work are −10 and −8 respectively. Reproduced from Koperski et al., 2D Mater. 6, 015001 (2018). CC BY 3.0 license.

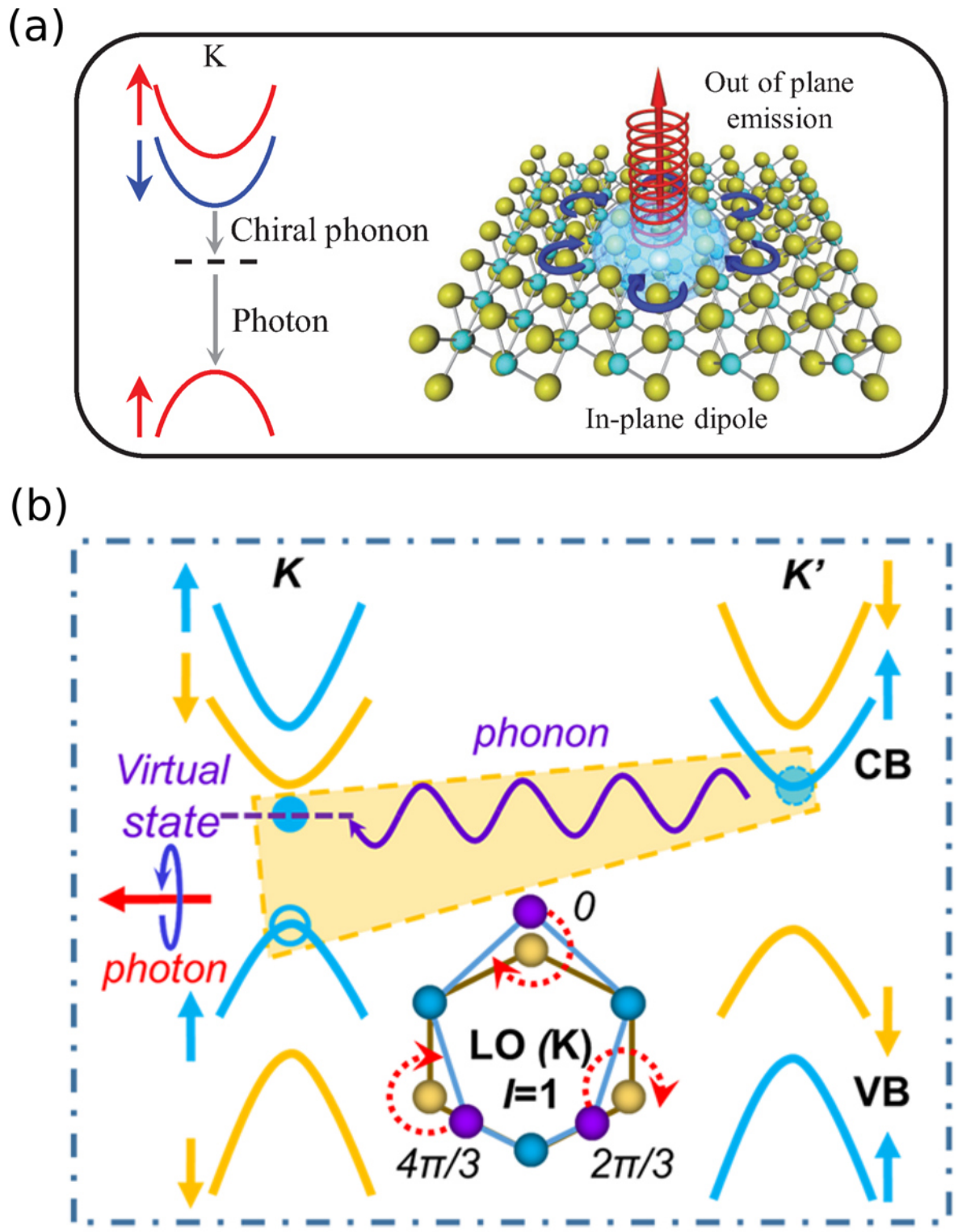

**Figure 9.** Possibility of a (a) spin-forbidden and (b) momentum-forbidden transition assisted by a chiral phonon. Transitions in (a) transitions show similar g factors in Faraday geometry as that of dark excitons and trions i.e. ~ − 9. A transition depicted in (b) shows a larger g factor ~ − 12. (a) is reproduced from Liu et al., Phys. Rev. Res. 1, 032007 (2019), Creative Commons Attribution 4.0 International license. (b) is reproduced with permission from Li et al., ACS Nano 13, 14107 (2019). Copyright 2019 American Chemical Society.

***2.1.2. Neutral 'dark' excitons in TMDCs.*** As shown in Figure 5, the dark excitons in TMDC monolayers exhibit a fine structure doublet due to an exchange interaction which mixes the valleys and leads to a splitting $\delta <$ 1 meV[170,181,206]. The low-energy state is dipole forbidden ($\Gamma_3$ symmetry), while the higher state ("gray" exciton with $\Gamma_4$ symmetry) has an out-of-plane dipole moment[170,206]. The gray excitons couple to the out-of-plane polarization of the light and their emission can be studied by collecting a PL component along the sample plane[168,170]. In a recent work, spin-forbidden dark excitons are resonantly controlled in a waveguide coupled monolayer $WSe_2$ device, providing access to the out-of-plane polarization[210]. Alternatively, these darkish states can be brightened by applying a magnetic field $B_\parallel$ in the plane of the sample (Voigt geometry, see Figure 6)[181,185,204–206,209,211]. The in-plane magnetic field mixes the bright and dark excitons through a Zeeman coupling of $g_\parallel \mu_B B_\parallel$ where $g_\parallel$ is the transverse (in-plane) g factor of the conduction electrons[204,205]. As a result, the dark excitons acquire some oscillator strength from the bright excitons leading to their emission in the out-of-plane direction[204,205]. $g_\parallel$ has been determined to be about −2 for $MoS_2$, $MoSe_2$ and $WSe_2$[182,185]. In another work, Yang et al. use time-resolved Kerr rotation under in-plane magnetic fields to determine the electronic g factor of $|g_0|$~1.86 in CVD-grown monolayer $MoS_2$[212]. Strikingly, this is very close to the free electron g factor and requires a theoretical understanding[30].

Magnetic-field-brightened dark and gray excitons are polarized along and perpendicular to the direction of the in-plane magnetic field, respectively[181,204]. The lifetime of dark excitons is longer than the bright excitons by about two to three orders of magnitude[170]. Therefore, they hold potential in investigating Bose-Einstein condensates and future

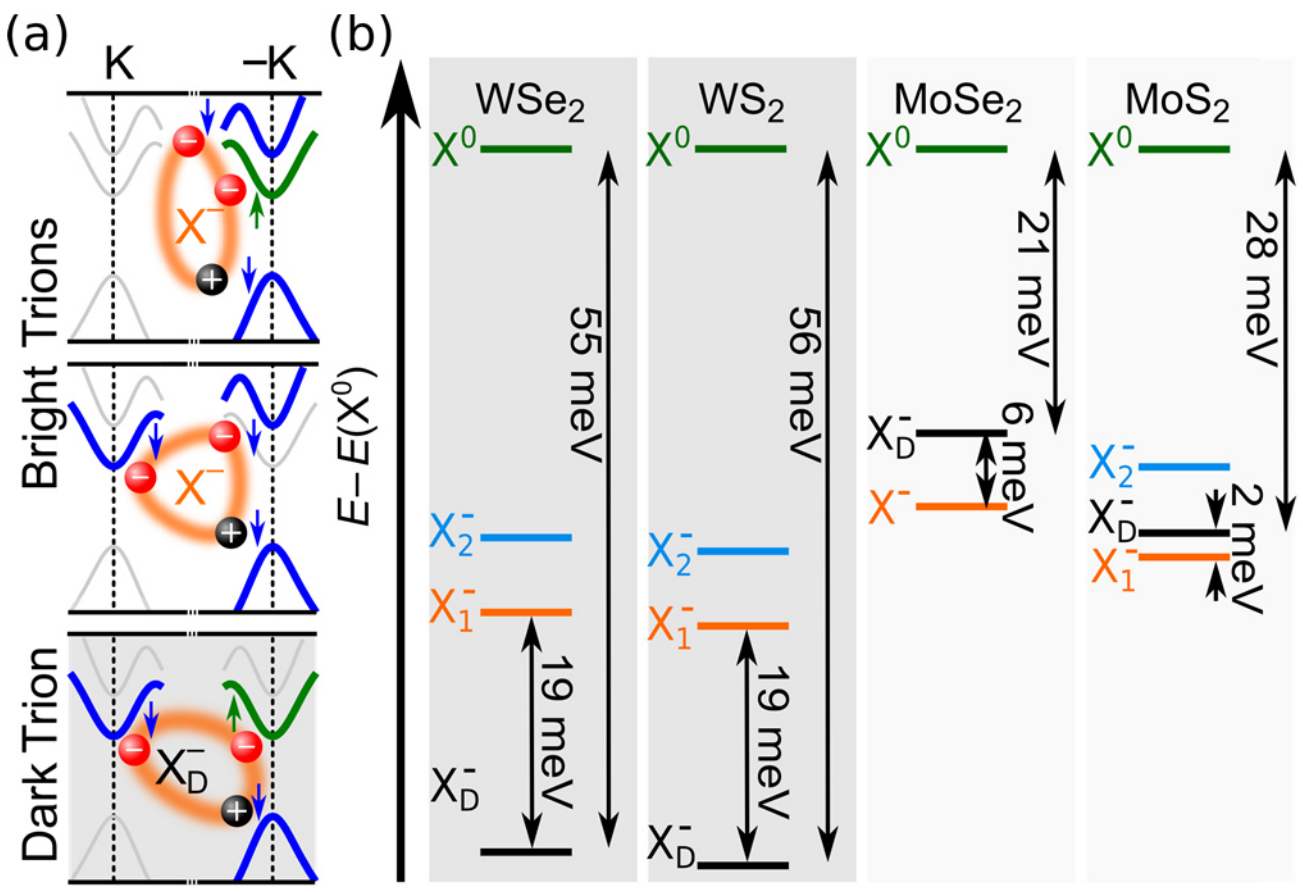

**Figure 10.** (a) Schematic showing conduction and valence bands involved in the creation of the bright ($X^-$) and the dark ($X_D^-$) trions in a doped monolayer TMDC. (b) In monolayer $WSe_2$ and $WS_2$, dark trions are the lower in energy than the two bright trions. In $MoSe_2$, the bright trion is the lowest in energy, while in $MoS_2$, the dark trion is almost coincident with the bright trion $X_1^-$. Figure adapted from Arora et al., Phys. Rev. B **101**, 241413 (2020). Copyright 2020 American Physical Society.

technological applications involving quantum information and computation using TMDCs.

To measure the magnetooptical response of dark excitons, two methods can be used. First, a large numerical aperture (NA) objective lens to collect PL emitted parallel to the sample plane can be used and magneto PL is performed in Faraday geometry[170]. Alternatively, if no PL from dark excitons using high NA objective is observed, one can apply tilted magnetic fields to the sample[181,182]. In this case, a component of the magnetic field within the plane of the sample brightens the dark excitons, and the out-of-plane component splits them. An example is shown in Figure 7.

Dark excitons are observed to exhibit larger g factors than the bright excitons[170,172–174,181,206]. The dark exciton g factors for the case of $WSe_2$, $MoSe_2$ and $MoS_2$ are found to be $-9$[170,172,174,206,208] (approximately), $-8.63$[182], and $-6.5$[182]. From the atomic picture considering contributions from the magnetic moments of the d-orbitals of the transition metal, the g factor of dark excitons is expected to be $-8$ which is close to the observed values (see Figure 8) [146]. For the case of $MoS_2$, a strong deviation of the dark exciton g factor from other materials is puzzling[182], and requires further theoretical and experimental work. This is similar to the bright exciton g factor of $MoS_2$ (discussed earlier) where a range of g factors in $MoS_2$ are found (between $-1.7$ and about $-4$) depending upon the exciton linewidth in the monolayer[182,201,202]. It needs to be checked if it is the case for the dark exciton in $MoS_2$ as well. Finally, phonon replicas of dark excitons, for instance in Figure 9 are observed energetically below the dark excitons in monolayer $WS_2$ and $WSe_2$ on the basis of their observed g factors (either about $-9$ or $-12$), energetic locations, and valley polarization[177,179,180,209]. Therefore, magnetooptical spectroscopy has been crucial in the identification of PL peaks previously under debate[87,166,203]. It needs to be understood if the emission lines exhibiting single-photon emission in TMDCs have their origin related to dark excitons/trions or their phonon replicas as well[213–217]. The author suspects this because the g factors of these emission lines are also large ($-9$ or $-13$ approximately)[214–217].

***2.1.3. Bright and dark trions.*** Trions are three-particle bound states with either two positive charges and a negative charge (positive trion), or two negative charges and a positive charge (negative trion). These quasiparticles have been extensively studied in doped quantum wells[57,58,218] and TMDCs[87,88,223–226,146,165,186,209,219–222] using optical spectroscopy and from theoretical perspectives. The binding energies of the trions in TMDCs are of the order of a few 10s of meVs, which is one to two orders of magnitude larger than the traditional quantum wells (typically 1 meV to few meVs)[218]. Therefore, it is possible to study trions in TMDCs at elevated temperatures. Similar to the neutral exciton, trions could be bright or dark depending upon the spin-forbidden optical selection rules[165]. We discuss bright trions first.

The initial state in the absorption process (or trion creation) is a free electron (hole) for the negative (positive) trion, while the final state is a trion[186,218]. In PL, the trion in its initial state relaxes to a free electron with the emission of a photon[186,218]. Depending upon the presence of the excess electron in the $K^+$ or $K^-$ valleys, three types of bright trions are possible[165]. Two of the bright trions and one dark trion are shown in Figure 10. Normally, these two bright trions are observed in the PL and absorption spectra of n-doped $MoS_2$[186,227], $WS_2$[105,226,228,229] and $WSe_2$[170,186], while one has been seen in $MoSe_2$ so far[185,186]. This is an ongoing matter of debate in the community where three close-lying trion resonances could be expected in $MoSe_2$ monolayer as well[165]. Furthermore, a discussion concerning the identification of trions (intravalley or intervalley) in the absorption and PL spectra of TMDCs has not been settled yet[187,228,230].

Both magnetoabsorption (reflectance)[146] and PL[99–105,139,187] have been used to determine the Zeeman splitting of the trions. In principle, in the low-density limit, the g factor of the trion is expected to be similar to that of the neutral exciton, which has been the case in most of the works[99,101–105,146,187]. This is because the magnetic moment of the extra electron contributes equally to the initial and final states in the absorption and PL processes. However, at larger carrier densities ($> 10^{12}$ cm$^{-2}$), many-body effects (such as excitons in a Fermi sea of electrons, or ‘Fermi polarons’) could change the magnetooptical response of the trions considerably, leading to a significant deviation from a value of $-4$[124,187,227,231]. A theoretical treatment (*ab initio* or parameter-based theories) explaining the variation of the g factor of trions due to these many-body effects is required to completely understand these phenomena. As such, the magnetoabsorption[146] and magneto PL[189] of bright trions have proven extremely useful to provide information on the g factors of the conduction and the valence bands forming the A excitons (and trions) of $WS_2$[146], $WSe_2$[189] and $MoSe_2$[146]. This is an important step towards developing a complete understanding of the spin-valley band structure of the TMDCs. However, many questions remain open. For instance, the g factors of the higher energy conduction bands were not determined in Ref. 146. Furthermore, the spin-orbit-split valence band which is involved in the formation of B exciton has not been investigated. In the future, the bright trions corresponding to the B excitons could be magnetooptically investigated to deduce this information.

Similar to dark excitons, magnetooptical spectroscopy has played a central role to identify and to investigate dark trions and their phonon replicas in monolayers of $WS_2$ and $WSe_2$[176–178,181,189,205,207,208]. An in-plane magnetic field can be used to brighten the dark trions (see Figure 11)[181,205]. Alternatively, a high NA objective lens may be used to capture a part of the in-plane emission from the dark trions[176–178,208]. The g factor of the dark trions are found to be around $-9$ or $-10$[176–178,181,189,205,207,208], on the similar order as that of the dark excitons discussed previously. This is close to the expected value of $-8$ from the atomic picture. On the other hand, a few works identify dark trion phonon replicas on the basis of their specific g factors (lying between $-9$ and $-13$), their emission energy positions with respect to the dark trions and their chirality using magneto PL spectroscopy (see Figure 10)[177,179,207,208]. The author is aware of one report where effective mass of the lower conduction band in hBN-encapsulated $WSe_2$ monolayer is found to be $m_e^* = 0.46m_0$ using magneto PL of intra- and intervalley trion emission lines[207].

***2.1.4. Biexcitons and charged biexcitons.*** Magneto-PL spectroscopy and the Zeeman effects have been useful in identifying the multiparticle complexes such as biexcitons and charged biexcitons in TMDC monolayers[172–174,232–234]. These investigations point out that a biexciton could possibly be considered as a composite of a bright and a dark exciton[172–175,211,232,233]. Similarly, a charged biexciton is reported to form from a bright trion and a dark exciton[172–175,233]. These quasiparticles are expected to have a g factor close to that of the neutral exciton and trion (i.e. about $-4$ for TMDCs)[211,235], which has been observed[172–175,232,233]. However, a magnetooptical signature of these many-body species is still lacking in Mo-based TMDCs and could be observed in future experiments[236].

The identification of dark excitons and trions, their phonon replicas, and other multiparticle complexes has settled some debate in the identification of the low-lying PL features in monolayer $WS_2$ and $WSe_2$. The question remains why is it different in the case of Mo-based TMDCs where no such low-energy peaks are observed in the PL spectra. A possible answer could emerge by comparing the dark-exciton occupation (expected to be low in monolayer $MoSe_2$ and $MoTe_2$ at low temperatures), exciton-phonon coupling strengths, and two energy scales in various TMDCs: i) the magnitude and sign of the conduction-band splitting, and ii) the optical-phonon energies which participate in the intravalley and intervalley scattering processes[237]. More theoretical inputs would be required to satisfactorily answer these questions.

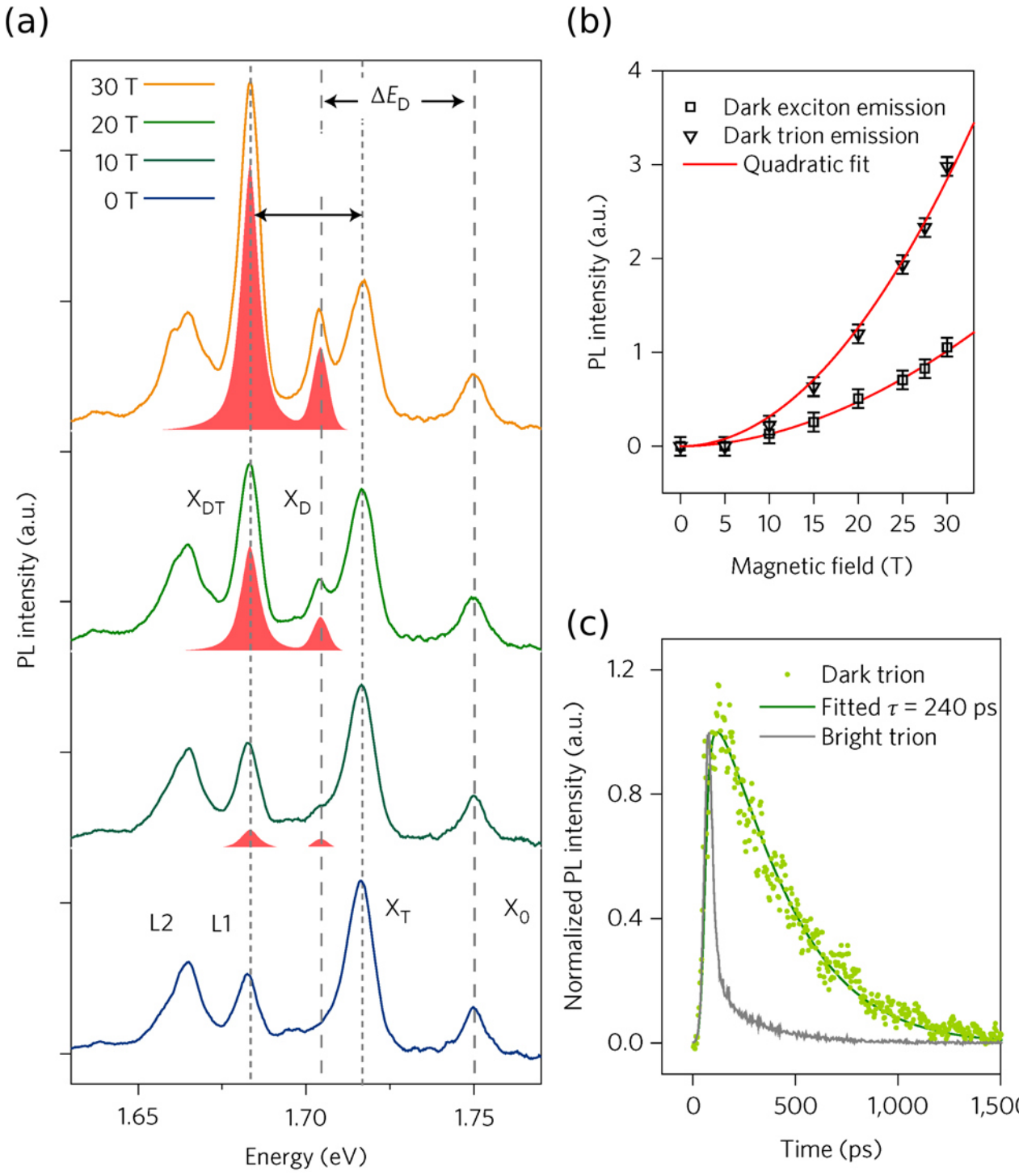

**Figure 11.** (a) Magnetic brightening of dark exciton and dark trion in monolayer $WSe_2$ demonstrated using magneto-PL in the presence of an in-plane magnetic field for various field strengths at a temperature of $T = 30$ K. (b) PL intensity of the dark exciton and trion with an in-plane magnetic field. Both features rise quadratically in intensity. (c) Time-resolved PL of the bright and the dark trions under an in-plane magnetic field $B_{\parallel} = 17.5$ T at $T = 4$ K. The dark trion decays much slowly than the bright trion. Reproduced with permission from from Zhang et al., Nat. Nanotechnol. 12, 883 (2017). Copyright 2017 Springer Nature.

***2.1.5. Interlayer excitons in homolayers.*** 'Interlayer excitons' with spatially-separated electron-hole wavefunctions have long lifetimes, and are promising for their role in investigating exciton Bose-Einstein condensates (BEC)[238]. They have a unique magnetooptical response in few-layer and bulk TMDCs[143,144,239]. The adjacent layers of these 'homolayers' with 2H stacking are oriented 180° with respect to each other as shown in Fig. 12(a). This results in a reversal of the magnetic moments of the corresponding bands (conduction and valence bands) in the neighboring layers, within a valley (Figure 12(d)). While the g factor of the A exciton in a monolayer or 'intralayer' exciton (exciton wavefunction largely located within a layer[143]) in few-layer or bulk TMDCs is of a negative sign, the g factor of interlayer exciton has a positive sign as shown in Figure 12. This was used as a fingerprint in the discovery of interlayer excitons in the reflectance spectrum of a 2H-stacked bulk TMDC (g factor of $+4 \pm 0.5$ in bulk $MoTe_2$, see Figure 12(b))[143], and later on in a bilayer and trilayer 2H-$MoS_2$[239]. Such positive g factors in Mo-based bulk TMDCs had been observed in 1970s[240], but their origin remained unexplained until recently. Magneto-reflectance spectra of a bilayer $MoS_2$ measured by us are presented in Figure 12(c) where an interlayer exciton g factor of $+7.8 \pm 0.3$ is obtained. We get a similar g factor for bulk $MoS_2$ as well (data not shown). This is close to an expected value of about $+8$ from a simple atomic picture in Figure 12(d). While resonances with positive g factors are found in $MoS_2$, $MoSe_2$, and $MoTe_2$, they are absent in W-based TMDCs[144]. The dissimilarity in the absolute value of the interlayer exciton g factor in bulk $MoTe_2$ and $MoS_2$ is a matter of debate. It is worth noting that in 3R stacked $MoS_2$ bilayers with a relative twist angle of 0° (compared to 180° in 2H bilayers), the interlayer hole tunneling is symmetry disallowed and interlayer excitons are not observed[241].

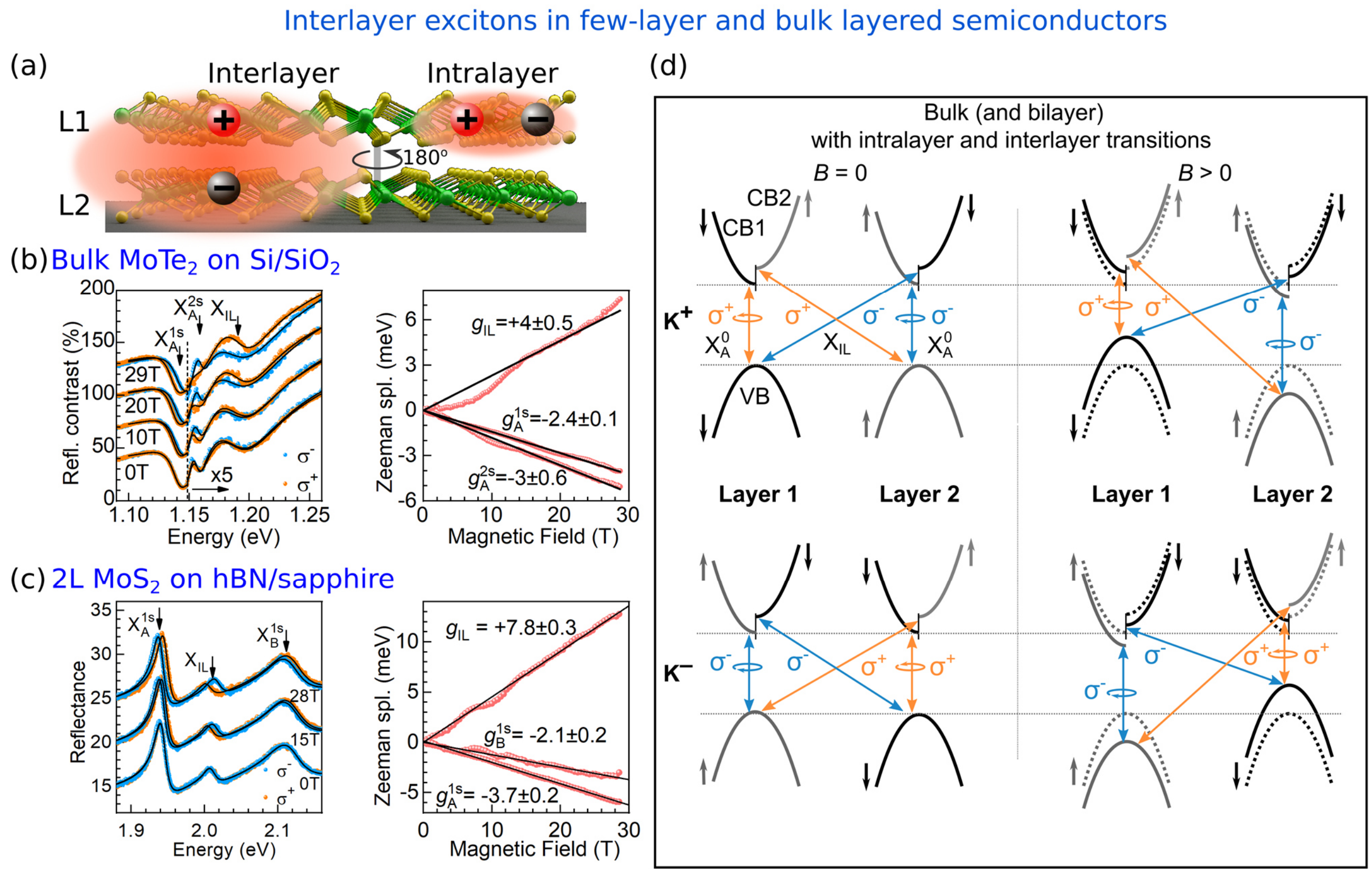

**Figure 12.** (a) A representation of intralayer and interlayer excitons in a bilayer TMDC (homolayer) with adjacent layers denoted as L1 and L2. (b) Helicity-resolved magneto-reflectance-contrast spectroscopy of intralayer and interlayer A excitons in a bulk $MoTe_2$ at $T = 4$ K[143]. Effective exciton g factors of intralayer and interlayer excitons are negative and positive, respectively. (c) Magneto-reflectance spectroscopy of a bilayer $MoS_2$ with an interlayer exciton g factor of $+7.8 \pm 0.3$. (d) Transition selection rules in the $K^+$ (upper panel) and $K^-$ (lower panel) valleys of a bilayer or a bulk TMDC. Layer 1 and Layer 2 denote adjacent layers as shown in (a). Left and right sides of the figure represent the movement of bands in the absence and presence of an out-of-plane applied magnetic field. Interlayer excitons have a positive g factor ($\sim +8$) compared to the negative g factors of intralayer excitons, in accordance with this simplistic picture. (b) and (d) are adapted from Arora et al., Nat. Commun. **8**, 639 (2017) (Creative Commons Attribution 4.0 International License)

*Ab initio* theory-based calculations showed that a relatively low spin-orbit coupling in Mo-based materials was responsible for a mixing between the interlayer excitons and the B excitons, providing a large oscillator strength to the interlayer excitons[144,242,243]. This mixing is negligible in W-based materials, leading to an unmeasurable interlayer exciton resonance in their reflectance spectra. Furthermore, an expected proximity of the interlayer exciton with the intralayer A exciton in W-based TMDCs also prohibits its detection[144]. It has been recently found that the oscillator strength of interlayer excitons in bilayer $MoS_2$ can be increased using an external electric field[244,245]. The question remains if the interlayer excitons in W-based TMDCs could be magnetooptically detected by enhancing their oscillator strength (e.g. using an electric field). Such long-lived dark interlayer excitons, if created as a lowest-energy state in a bilayer system, such as by applying a strain, could hold potential in studying collective physics such as BEC.

***2.1.6. Interlayer excitons in heterobilayers.*** So far, we have discussed magnetooptics of monolayer and few-layer TMDC homostructures. However, following the traditional heritage of III-V and II-VI heterostructures, research is progressing at a rapid pace in the area of heterostructures of layered materials[106,246–252]. Stacks of layered crystals can be readily created by stamping monolayers or few layers of TMDCs on the top of each other[246]. In a heterobilayer, hexagonal lattice units of the adjacent layers can be twisted with respect to each other at an angle $0° \leq \theta \leq 60°$ and the electronic properties could be tweaked. This emerging area of research is termed 'twistronics'[253]. For a slightly different lattice constant and alignment angle between the stacked layers, a moiré pattern is formed as shown in Figure 13(d)[254]. For $\theta = 0°$ and $\theta = 60°$, AA and AB stacked configurations are created, respectively[106,247–249]. An example of an AB stacked $MoSe_2$-$WSe_2$ heterobilayer is shown in Figure 13(a). One notices that the $K^{\pm}$ valleys of $MoSe_2$ align with $K^{\mp}$ valleys of $WSe_2$. The interlayer excitons created between the type-II band alignment (Figure 13(b)) in this structure is lower in energy than the A excitons in the individual layers. Due to the specific valley alignment, the valley magnetic moment contributions enhance the g factor ($\sim -15$ which corresponds to a Zeeman splitting of about 26 meV at 30 T) of interlayer excitons significantly compared to the case of monolayers (Figure 14 (c))[106]. In the case of AA stacked layers, a doublet with a spin-conserving and spin-flipping transition is observed, with g factors of $-8.5 \pm 1.5$ and $+7.1 \pm 1.6$ respectively[247]. It has been shown in multiple theoretical works that such spin-flipping transitions may be brightened in heterobilayers with matrix elements similar to those of spin-conserving transitions, due to the presence of a periodic moiré potential (Figure 13(d) and (e))[255,256]. A few other recent works have also observed interlayer excitons in moiré potential with large negative ($\sim -16$) and positive g factor ($\sim +7$)[248,250–252]. The role of chiral-phonon-induced exciton-phonon interactions in the $WSe_2$/$MoSe_2$ heterostructures in the determination of the optical selection rules and brightening of spin-flip (triplet) transitions has been highlighted[251]. The area of research is still in its infancy, and in the future, heterobilayers of other semiconducting TMDCs are expected to be investigated magnetooptically. Many parameters such as twist angle, layer thickness, doping, electric fields etc. could be tuned to create heterostructures exhibiting interesting physics and potential for future valleytronics and twistronics device-based applications.

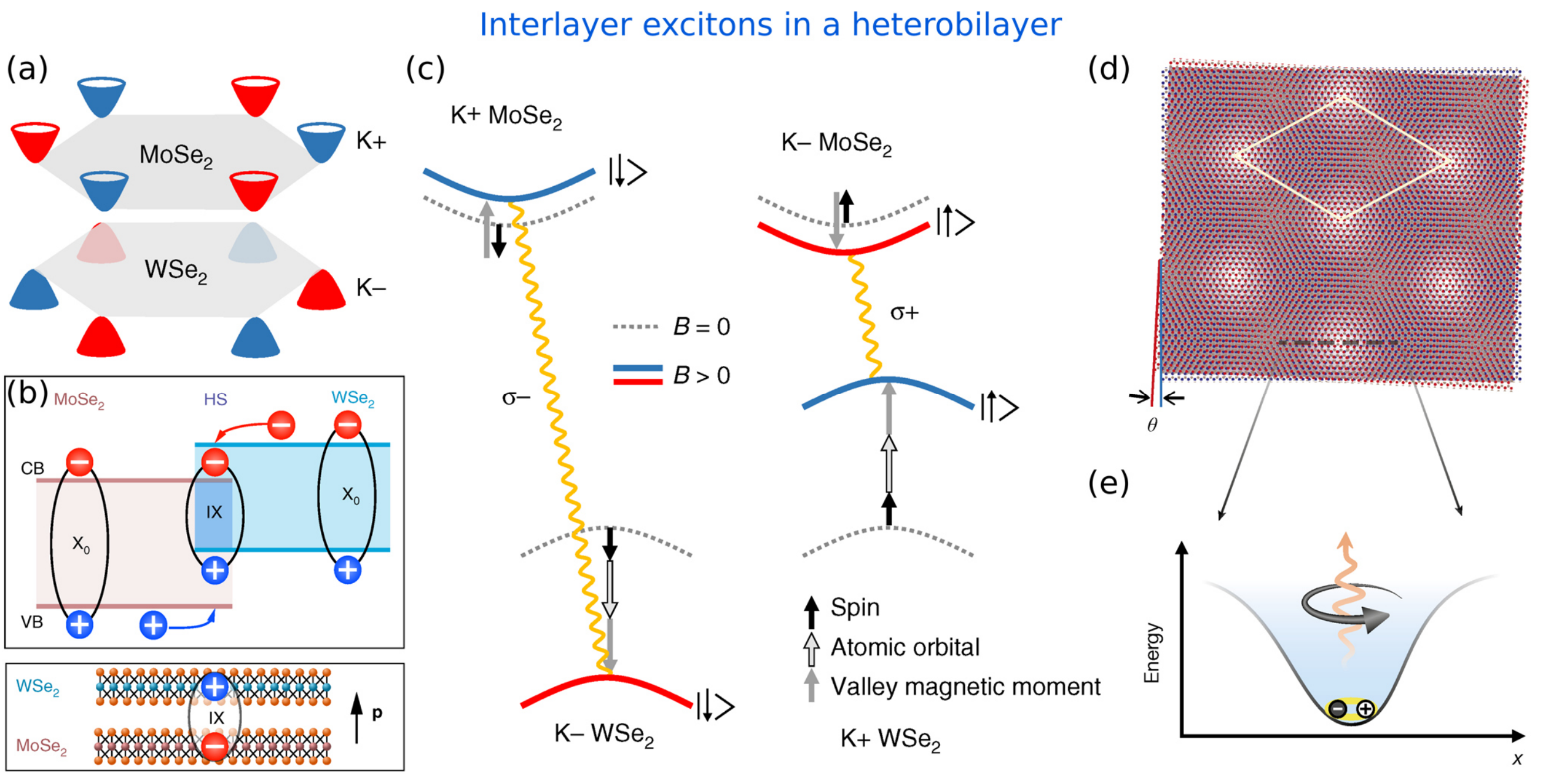

**Figure 13.** (a) Alignment of $K^{\pm}$ valleys in $MoSe_2$-$WSe_2$ heterostructure with AB stacking (From Ref. 106). (b) Type-II band alignment and creation of the interlayer exciton in this heterostructure (From Ref. 247). (c) Sketch of circularly-polarized interlayer exciton transitions in AB stacked heterostructure in (a) in the absence (dashed lines) and presence (solid lines) of an out-of-plane magnetic field. Characteristic alignment of $K^{\pm}$ valleys between two layers lead to a large interlayer exciton g factor (From Ref. 106). (d) Moiré superlattice formed in a heterobilayer with a small twist angle $\theta$, and (e) exciton in a moiré potential (From Ref. 248). (a) and (c) are reproduced with permission from Nagler et al., Nat. Commun. 8, 1551 (2017), Creative Commons Attribution 4.0 International License. (b) is reproduced with permission from Ciarrocchi et al., Nat. Photonics 13, 131 (2019), Copyright 2019 Springer Nature. (d) and (e) are reproduced with permission from Seyler et al., Nature 567, 66 (2019), Copyright 2019 Springer Nature.

## 2.2 Valley polarization and coherence in magnetic fields.

When the valley degeneracy is lifted, a magnetic field leads to a redistribution of the carriers between the valleys. This leads to an uneven intensity of the emission of the two circular polarizations, and is termed as magnetic-field-induced 'valley polarization'[99–106]. An example is shown in Figure 3(a) and (b). In Ref. 146, the authors assumed a Boltzmann distribution of carriers (low-doping-density regime) in the conduction bands of doped monolayers of $WS_2$ and $MoSe_2$, and studied the valley polarization of trions in the magnetic field. It was used to determine the lower conduction band g factor equal to 1.08 and 1.84 for $WS_2$ and $MoSe_2$, respectively. The magneto-reflectance contrast of trions in monolayer $MoSe_2$ is shown in Figure 14 where the trions appear to be completely polarized around 10 T. Magnetic-field-induced valley polarization of trions in $WS_2$ is much more complicated. Two bright trions (singlet and triplet, shown in Figure 10(a)) are observed in reflectance and PL and show different polarization response under magnetic fields[146,230]. In reflectance, the polarization is of opposite sign and similar magnitude[146,230]. This is explained by an uneven occupation of states in the $K^{\pm}$ valleys by free carriers due to valley Zeeman splitting, which affects the oscillator strengths in this peculiar way. However, in magneto PL, only the singlet trion is significantly polarized[230]. This has not been explained satisfactorily and could involve an interplay between many processes such as intervalley relaxation of carriers, spin-flip processes, and laser-induced photodoping effects. Magneto PL as a function of gate voltage (carrier density) and excitation intensity is expected to shine light on this problem in the future.

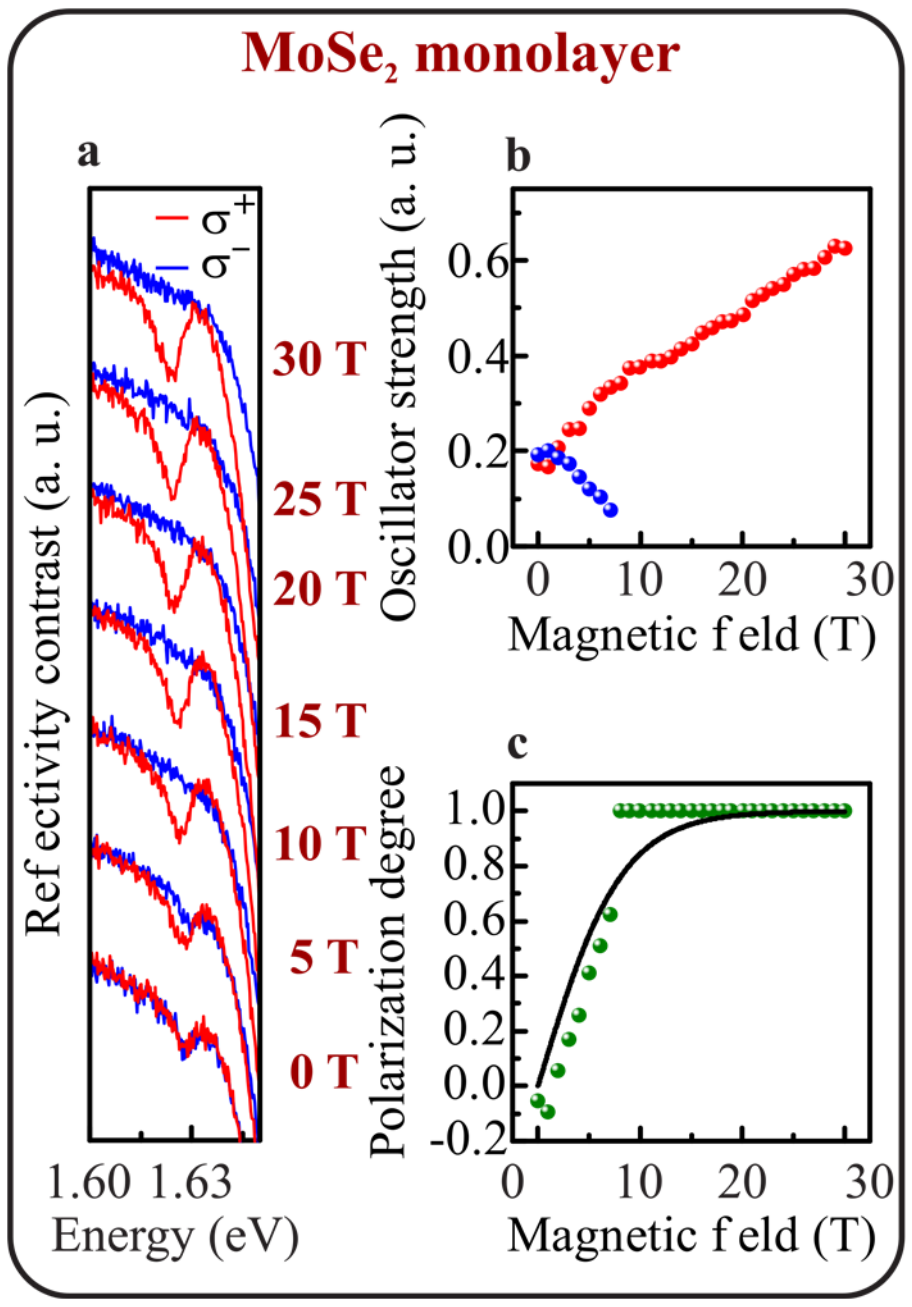

**Figure 14.** (a) Magneto-reflectance contrast spectra of trions in an unintentionally doped monolayer of $MoSe_2$ under magnetic fields of up to 30 T for the two circular polarizations. Trions polarize strongly in the presence of magnetic field due to a redistribution of carriers between the energetically non-degenerate valleys. The oscillator strengths of the trions are shown in (b) while (c) depicts the degree of polarization. The data has been fitted with a model based on Boltzmann distribution of carriers and conduction band g factor is determined equal to 1.84. Reproduced with permission from Koperski et al., 2D Mater. 6, 015001 (2018). CC BY 3.0 license.

It is worth discussing the peculiar case of doped $MoS_2$ monolayer under magnetic fields. Roch et al. performed helicity-resolved magneto-reflectance on this system as a function of carrier density[227,257]. The unusual trion polarization under magnetic fields in the low-density regime ($n \leq 5 \times 10^{12}\text{cm}^{-2}$) seems to suggest that the ground-state is spin polarized. At a certain critical density ($n_c = 3.0 \times 10^{12}\text{cm}^{-2}$ in Ref. 257), the system is reported to undergo a ferromagnetic-to-paramagnetic first-order phase transition[257]. The mechanism behind the spontaneous spin polarization is assumed to be the strong intra- and inter-valley exchange interaction shown in Figure 15. However, a microscopic theoretical description of this effect in $MoS_2$ requires to be developed for a complete picture. This is a developing area of research, where alternative views are emerging[187]. From an experimental perspective, the possible future directions could be to test the spin-polarization effects by placing monolayer $MoS_2$ on magnetic substrates, or on 2D layered magnetic materials. It might be possible to achieve significant spin polarization under low magnetic fields (compared to the high fields $B_z =$ 9 T) in Ref. 257.

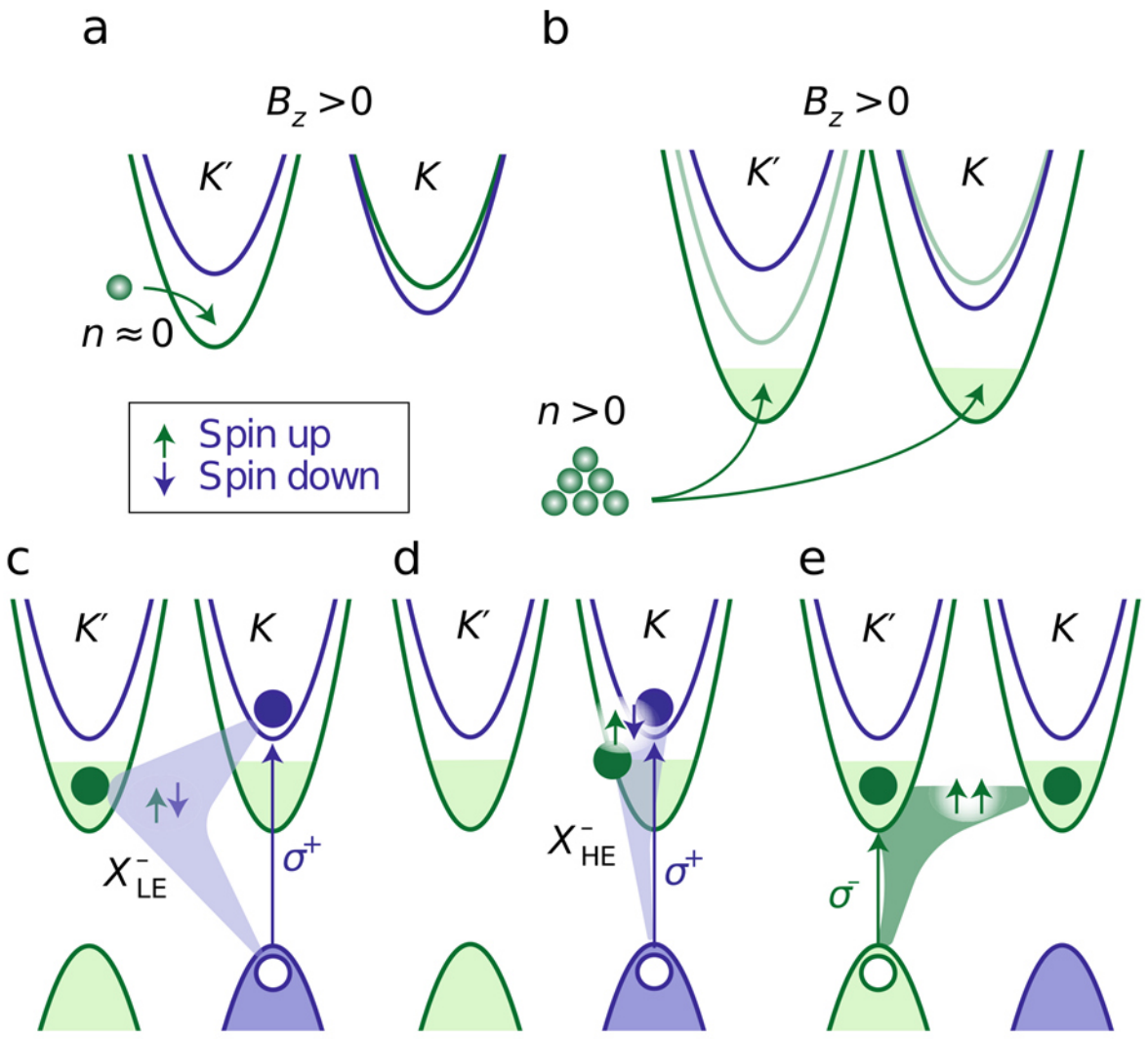

**Figure 15.** Schematic depicting the proposed explanation of observed valley polarization of trions under an out-of-plane magnetic field in the case of a doped $MoS_2$ monolayer[227]. (a) represents the valley Zeeman effect in the conduction bands under magnetic field $B_z$. The first electrons (green spheres) enter the lowest energy band. (b) As the electron density is raised, the intra- and inter-valley exchange interactions pull the same-spin conduction bands to low energy, resulting in a spin-polarized 2D electron gas. (c) and (d) depict the low- and high-energy exciton polarons $X^-_{LE}$ and $X^-_{HE}$ respectively. The corresponding trions in both cases are spin singlets. (e) The spin triplet trion is unbound which results in a high-energy tail of the neutral exciton of $\sigma^-$ polarization. Reproduced with permission from Roch et al., Nat. Nanotechnol. 14, 432 (2019). Copyright 2019 Springer Nature.

Linearly-polarized light can be written as a superposition of two circular polarizations of opposite orientations. Therefore, both $K^+$ and $K^-$ valleys can be excited simultaneously using linearly-polarized light, and a coherent superposition of states can be created in the two valleys ('valley coherence')[93–98]. TMDCs such as monolayer $WS_2$ and $WSe_2$ when excited using linearly-polarized light also emit a linear polarization, while the polarization angle of the emission follows that of the excitation[93–98]. An external magnetic field leads to a Zeeman splitting between the two valleys, lifting the valley degeneracy. This leads to a phase difference between the two circularly-polarized components. As a result, the emission polarization rotates with respect to the excitation polarization as shown in Figure 16. The coherent state decays with a lifetime $\tau_c$ (or $t_2^*$) which is of the order of a few 100 fs at liquid He temperatures[96–98,258]. This decay leads to a depolarization of the emission polarization. The extent of this depolarization increases as the external magnetic field rises[96–98]. Therefore, external magnetic fields have helped in understanding valley-coherence effects, deducing the coherence lifetimes, and controlling the intervalley phase. These developments mark important steps towards the realization of an excitonic valleytronic gate for future quantum technologies such as in quantum computation. However, at the moment, many roadblocks persist. The coherence life times are small, and the magnetic fields required for such a phase control are much higher (up to 25 T external fields were used[96,97]) than suitable for practical device purposes. Additionally, the experiments were done at cryogenic temperatures, while investigations remain to be performed at room temperature.

On the contrary, a long spin coherence lifetime of resident carriers exceeding many nanoseconds to a few microseconds is observed in monolayer TMDCs using time-resolved Kerr rotation spectroscopy[212,259–266]. This is many orders of magnitude larger than the PL decay time and the excitonic valley coherence decay time discussed above. In fact, while the electron polarization decay (nanosecond regime) was seen to accelerate in the presence of an in-plane magnetic field (Hanle effect)[212,259,262,263], the hole polarization in $WSe_2$ monolayer persisted for a couple of microseconds, and did not depend upon the magnetic field[262]. This verified a strong spin-valley locking of holes in the monolayers. Therefore, the application of a robust spin coherence in the future monolayer TMDC-based spintronic devices is promising. However, so far, this long spin coherence persists only at cryogenic temperatures which might limit usage in device applications, and requires further work.

Recently, there are proposals to optically control the entanglement of electron and nuclear spins in TMDC monolayers via intervalley terms of the hyperfine interaction[267–269]. It has been argued that similar to the case of electron and hole hyperfine interactions in III-V semiconductors[270,271], they are of short range nature in TMDCs as well. Such a short-range hyperfine interaction is expected to lead to a coupling between electronic states in the $K^{\pm}$ valleys providing a spin relaxation mechanism. On the other hand, within a valley of TMDC, the electron-nuclear spin flip is suppressed due to a large (4 orders of magnitude larger than hyperfine splitting) spin-orbit splitting[267]. Further understanding of the role of hyperfine interaction in spin coherence dephasing is expected from future experimental and theoretical works.

**2.3 Landau quantization and Rydberg spectroscopy: exciton binding energy, reduced masses, and oscillator strength enhancement.** The problem of a 3D and a 2D electron-hole system under strong magnetic fields has been extensively studied theoretically[38,40–44,50–52,272]. The conduction and valence bands are quantized in the presence

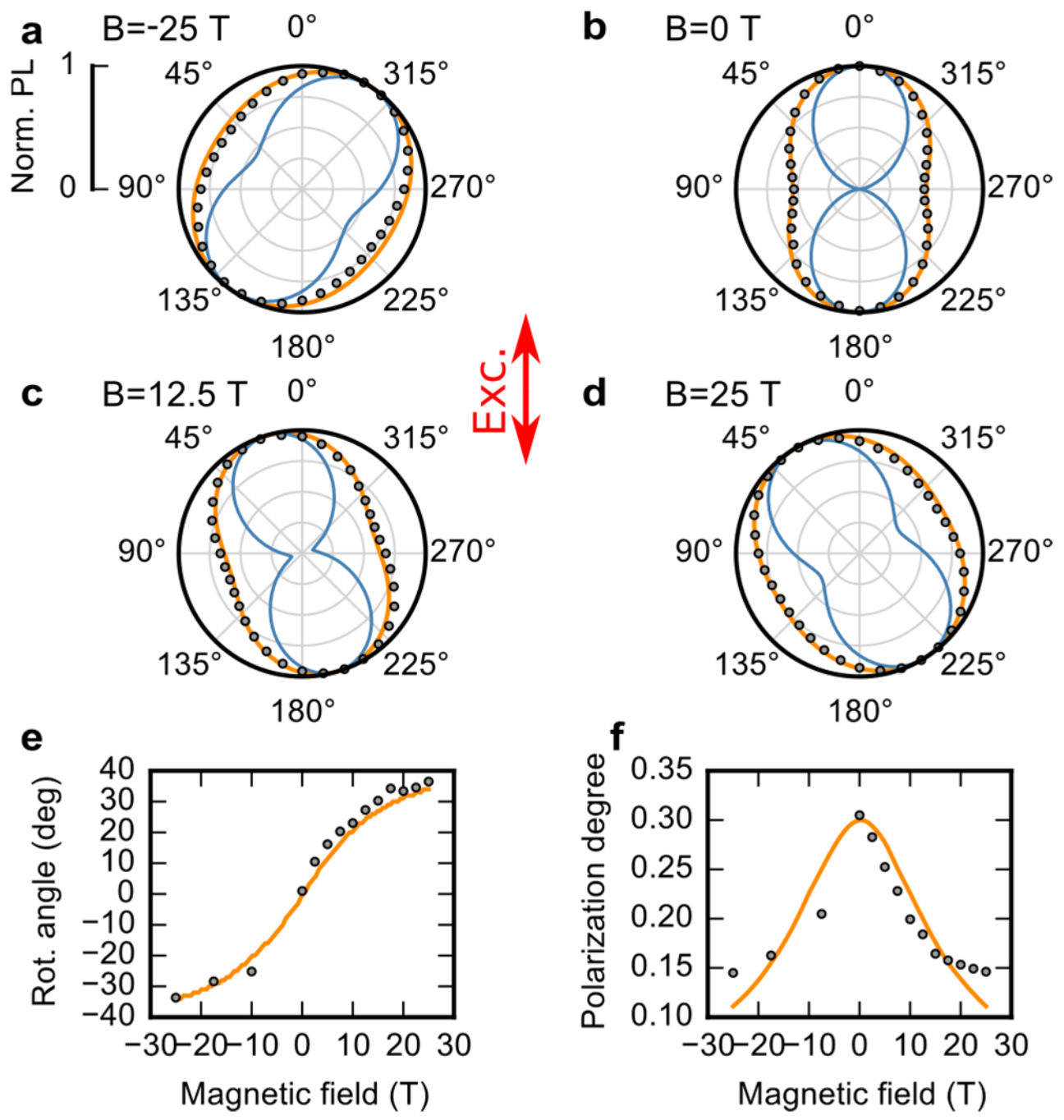

**Figure 16.** Linear polarization of luminescence from coherently emitting valleys can be manipulated using an out-of-plane externally applied magnetic field. An example of the magnetic-field-induced rotation and depolarization of emission polarization from a monolayer $WS_2$ under a linearly-polarized excitation along 0° (red double-headed arrow) is shown. (a), (c) and (d) are the emission-polarization intensities as a function of angle under magnetic fields of −25 T, −12.5 T and 25 T, respectively. (e) and (f) are the rotation angle and the polarization degree as a function of magnetic field. The data is used to determine the decay time of the coherent light emission i.e. $t_2^* = 260$ fs. Reproduced from Schmidt et al., Phys. Rev. Lett. 117, 077402 (2016). Copyright 2016 American Physical Society.

of a magnetic field (Landau quantization of the cyclotron orbits), and the effective dimensionality of the system is reduced by two.[2] Energy of the $N^{th}$ Landau level of the conduction ($c$) or the valence ($v$) band is given by

$$E_{LL}^{N(c,v)} = \left(N^{c,v} - \frac{1}{2}\right)\hbar\omega^{c,v} \qquad (3)$$

where $\omega^{c,v} = eB/m_{c,v}^*$ is the cyclotron frequency of the conduction or valence band, and $m_{c,v}^*$ is the effective mass of the band. The $N^{th}$ inter-Landau-level transition follows the selection rule $\Delta N = N^c - N^v = 0$.[4] In the presence of the Coulomb interaction, the $N$th inter-Landau transition corresponds to optically probing the $N$s exciton (magnetoexciton) state[38,52,53,55,56,129–131,202]. Apart from separating the excited exciton states from each other, a magnetic field also enhances their oscillator strengths linear with $B$, by squeezing the exciton wave functions[38,51,272]. This is especially advantageous for performing Rydberg spectroscopy where Rydberg states with very large $N$ could be observed under magnetic fields[45,47,55,56,129–131,202]. Recently, Rydberg states up to 11s have been observed in hBN-encapsulated monolayer $WSe_2$ in moderately large magnetic fields of 17 T[131] (see Figure 17 (e) and (f)).

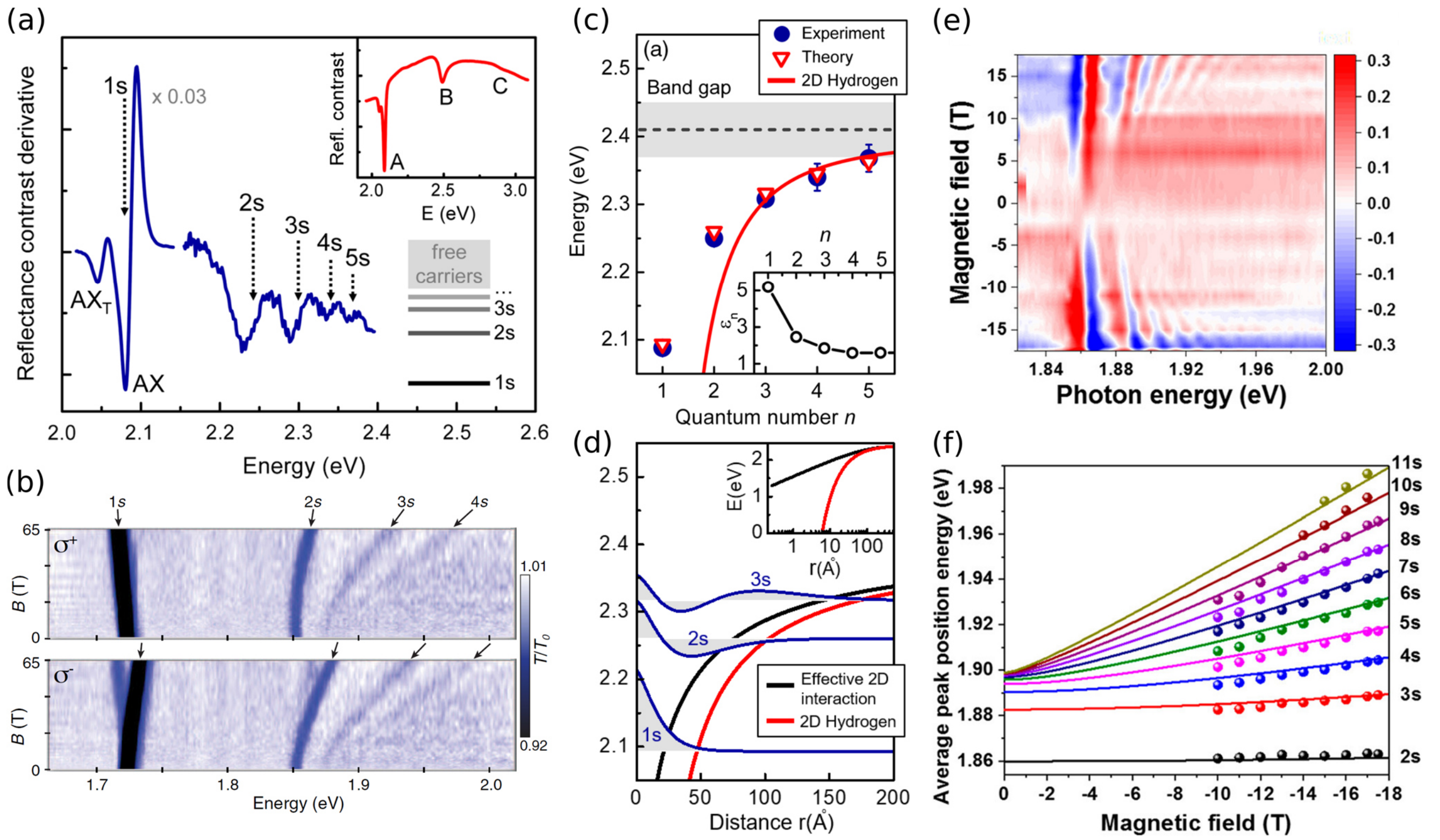

**Figure 17.** (a) Rydberg states of excitons up to $n = 5$ in monolayer $WS_2$ measured using reflectance contrast spectroscopy in the absence of magnetic field. (b) Circular-polarization-resolved magnetotransmittance of Rydberg excitons up to $n = 4$ in hBN-encapsulated monolayer $WSe_2$ under pulsed high magnetic fields of up to 65 T. 1s and 2s show a quadratic diamagnetic shift throughout the field range, suggesting a low-field limit. 4s energy blue shifts approximately linear with $B$ at large fields, representing a crossover to the high-field limit. (c) Experimental and theoretical $n$s exciton transition energies from (a), along with a 2D hydrogen model (red solid line) for comparison. The inset shows the corresponding effective dielectric constant. (d) Screened 2D interaction potential (black line, Eq. 4) and 2D hydrogen potential (red) along with the corresponding energy levels and radial wave functions up to $n = 3$ for data in (c). The inset shows the potentials in a semilogarithmic plot. (e) Valley polarization (magnitudes marked by color scale in the legend) and (f) energy positions (Landau-fan diagram) of Rydberg excitons up to $n = 11$ deduced from the magnetophotocurrent spectroscopy of hBN-encapsulated $WSe_2$ monolayer, with solid-line fits using a numerical model based on Rytova-Keldysh potential. Valley polarization is defined as $(PC_{K+} - PC_{K-})/(PC_{K+} + PC_{K-})$. A linear blue shift of 9s to 11s excitons for $B > 10$ T points towards a high-field limit of these states. The A exciton binding energy determined from the fan diagram is 168.6 meV. (a), (c) and (d) Reproduced with permission from Chernikov et al., Phys. Rev. Lett., 113, 076802 (2014). Copyright 2014 American Physical Society. (b) Reproduced with permission from Stier et al., Phys. Rev. Lett. 120, 057405 (2018). Copyright 2018 American Physical Society. (e) and (f) Reproduced with permission from Wang et al., Nano Lett. 10, 7635 (2020). Copyright 2020 American Chemical Society.

Due to the excitonic effects, one does not observe a simple linear $B$-dependent blue shift of the exciton transition energy, especially for low $N$ inter-Landau transitions. An example is shown in Figure 17(b) for exciton states up to $n = 4$ in hBN-encapsulated $WSe_2$[129]. We clearly observe a non-linear behavior of 1s and 2s exciton energies even under high magnetic fields of up to 65 T. However, 3s and 4s states begin to crossover to a linear blue shift regime for higher fields (above about 50 T). To understand this, we compare the Coulomb energy scale given by $E_b^N = Ry^*/(N - 1/2)^2$ (where $Ry^*$ is the effective Rydberg) with the ground-state cyclotron energy $\hbar\omega^{c,v}/2$. We define a dimensionless parameter $\gamma = \left(\frac{\hbar\omega^{c,v}}{2}\right)/E_b^N$.[2] The parameter $\gamma$ is also written in terms of a magnetic length $l = \sqrt{\frac{\hbar}{eB}}$ which depends only on magnetic field, and the Bohr radius $a_B^*$ as $\gamma = \left(\frac{a_B^*}{l}\right)^2$.

Under low magnetic fields, $\gamma \ll 1$ i.e. the Coulomb energy dominates. Often, in the case for the ground-state exciton, $\gamma \ll 1$ until under very high fields. In this limit, only a $B^2$-dependent diamagnetic shift of the exciton is observed. This low-field limit is discussed in the next section. In high magnetic fields, when $\gamma \gg 1$, the system is purely Landau like. This high-field limit is readily approached under moderate fields for $Ns$ magnetoexcitons with a large $N$, while the exciton energy blue shifts linearly with magnetic field approximated by $\Delta E_{LL}^N$[2,129–131]. For instance, the binding energy of the $11s$ exciton in Figure 17(f) is 1.2 meV, which is much smaller than the Landau quantization energy of 8.69 meV at 15 T. Therefore, we are in the high-field limit. Another example of an hBN-encapsulated monolayer $WSe_2$ is shown in Figure 18.[130] Here, the 5s exciton nearly lies in moderately high-field limit ($\gamma$~1) above $B = 9$ T. The energies of the ground and excited exciton states are commonly plotted as a function of $B$ in a Landau-fan diagram with exciton energy and magnetic field as the **x-** and **y-**axis respectively such as in Figures 17(f) and 18(d)[2,4,45,47,55,56,129–131,202]. An analysis of this data has been readily used in the literature for a determination of the exciton-reduced masses $\mu_X^*$ and their binding energies $E_b$ in III-V and II-VI QWs[2,47,55] as well as in TMDCs[129–131,202]. We now discuss the parameters derived for TMDCs.

Fitting a nearly linear shift of 3s and 4s excitons at high magnetic fields in Figure 17(b) (moderately high-field limit i.e. $\gamma \sim 1$ or slightly larger), the authors determine an upper limit on the exciton reduced mass equal to $0.23\ m_0$[129] ($m_0$ is the free electron mass). Molas et al. use magneto PL of the $5s$ state and find this value $= 0.25\ m_0$[130]. In Figure 16(d), the high-field limit ($\gamma \gg 1$) of 9s, 10s and 11s Rydberg states in hBN-encapsulated $WSe_2$ monolayer is used to determine a reduced mass of $0.2\ m_0$[131]. Theory predicts a value of $0.17\ m_0$, deviations from which are a matter of debate[163,221]. Wang et al. measure inter-Landau Level transitions in a gated (for tuning the carrier density) hBN-encapsulated $WSe_2$ monolayer using magnetoreflectance spectroscopy when the electron-hole interaction could be neglected[273]. They find the reduced mass equal to $0.268\ m_0$. A few more recent works also reported inter-Landau-level transitions in doped hBN-encapsulated monolayer $WSe_2$[274–276] and $MoSe_2$[277]. Notably, the reduced masses are also determined in a more consistent fashion by analyzing the energy shifts of the excitons in the low-field limit (diamagnetic shift). We discuss this separately in the next section.

The Landau-fan diagram provides an estimate of the exciton binding energy. The extrapolation of the high $N$ Rydberg states to 0 magnetic field asymptotically approaches the free-particle band gap as shown in Figure 17(d). Subtracting the 1s exciton energy from this provides an estimate of the exciton binding energy. Popularly, a nonhydrogenic screened Rytova-Keldysh potential[278,279] has been used to explain the observed energy shifts of the excited exciton states in hBN-encapulated monolayers of TMDCs[129,130,202]

$$V(r) = -\frac{e^2}{8\epsilon_0 r_0}\left[H_0\left(\frac{\epsilon r}{r_0}\right) - Y_0\left(\frac{\epsilon r}{r_0}\right)\right] \qquad (4)$$

where $H_0$ and $Y_0$ are Struve and Bessel functions respectively, $\epsilon$ is the average dielectric constant of the top and the bottom layer, $\epsilon_0$ is the permittivity of the vacuum, $r_0 = 2\pi\chi_{2D}$ is the screening length and $\chi_{2D}$ is the 2D polarizability. Solid lines in Fig. 17(f) are the numerical fits to the data using this model where the best fit parameters are $\epsilon = 4.3, r_0 = 4.5$ nm and $\mu_X^* = 0.2\ m_0$.[131] The A exciton binding energies determined using this treatment in various works are mentioned in Table 1.

Therefore, magnetooptical spectroscopy has provided answers to some of the most fundamental questions concerning exciton band structure. Indeed, the binding energies and reduced masses of the excitons in the free-standing monolayers still requires to be determined magnetooptically. One would have to wait for high-quality free-standing samples where strain-free monolayers would be exfoliated or CVD grown on substrates with holes (at least 10 µm diameter to avoid diffraction effects in transmission mode) drilled through them. The *ab initio* theory calculations predict that the binding energies in this case could be of the order of 1 eV for monolayers[126,221,224], and 0.1 eV for bulk TMDCs[143,199]. Furthermore, the Rydberg spectroscopy of a few-layer and bulk TMDCs has not been performed so far for $N > 2$. To understand the variation of exciton-reduced masses with layer thickness, magneto-reflectance or transmittance measurements would be preferable over magneto PL. It is because thicker layers being indirect gap semiconductors are inefficient emitters of light around their direct band gaps[166,169,183,280–285].

**Table 1.** Reduced mass and binding energy of A excitons in TMDC monolayers determined from a combined analysis of the ground- and excited-exciton-state energies in the low- and high-field limits, along with the references.

| Material | Reduced A exciton mass $\mu_X^*$ ($m_0$) | A exciton binding energy (meV) |
|---|---|---|
| 1L $WS_2$ on $SiO_2$ | - | 410[142] |
| 1L hBN/$MoS_2$/hBN | 0.275[202], 0.26[130] | 221[202], 217[130] |
| 1L hBN/$MoSe_2$/hBN | 0.35[202], 0.44[130] | 231[202],216[130] |
| 1L hBN/$WS_2$/hBN | 0.175[202], 0.15[130] | 180[202], 174[130] |
| 1L hBN/$WSe_2$/hBN | 0.20[202], 0.21[130], 0.20[131] | 167[202], 167[129], 168.6[131], 167[130] |
| 1L hBN/$MoTe_2$/hBN | 0.36[202] | 177[202] |

**2.4 Diamagnetic shift: exciton masses and Bohr radii.** Magnetooptical measurements on excitons in the low-field limit ($\gamma \ll 1$) are useful in determining the exciton masses and Bohr radii[2,48]. The $B^2$ dependent diamagnetic shift of exciton energy in this limit is given by[2,38,129–131,142,202,286,287]

$$\Delta E_{dia} = \frac{e^2}{8\,\mu_X^*}\langle r^2\rangle B^2 \qquad (5)$$

where $a_{rms} = \sqrt{\langle r^2 \rangle}$ is the root-mean-square (RMS) radius of the exciton, and $\mu_X^*$ is the reduced mass. This is a general result, applicable equally to 3D and 2D ground and excited-state excitons. For the exciton ground ($1s$) state, the diamagnetic shift is given in 3D and 2D as[2,49]

$$\Delta E_{dia}^{3D}(1s) = \sigma B^2 = \frac{e^2 {a_{B,3D}^*}^2 B^2}{\mu_X^*},\ \sigma = \frac{4\pi^2\kappa^2\epsilon_0^2\hbar^4}{e^2{\mu_X^*}^3} \qquad (6)$$

$$\Delta E_{dia}^{2D}(1s) = \frac{3}{16}\sigma B^2 \qquad (7)$$

$\kappa$ is the dielectric constant of the material, and $a_{B,3D}^* = (4\pi\kappa\epsilon_0\hbar^2)/(\mu_X^* e^2)$ is the 3D exciton Bohr radius. The exciton Bohr radius is defined here as the radius at which the wavefunction has decayed to $1/e$ of its maximum[288]. It is worth noting that the 3D Bohr radius is twice as large than the 2D case $[a^*_{B,2D} = (2\pi\kappa\epsilon_0\hbar^2)/(\mu^*_X e^2)]$ and the 3D exciton diamagnetic shift is larger by a factor of 16/3 than in 2D. Also, the exciton RMS radius in 3D is $a_{rms,3D} = \sqrt{3}a^*_{B,3D}$ and in 2D is $a_{rms,2D} = \sqrt{\frac{3}{2}}a^*_{B,2D}$. The diamagnetic shifts of the excited states of the excitons are much larger than the ground state (large Bohr radii), and are discussed in Refs. 2,38,48,50. For the 3D exciton, the diamagnetic shifts of the excited states are given by

$$\Delta E^{3D}_{dia}(ns) = \frac{n^2(5n^2+1)}{6}\sigma B^2 \quad (8)$$

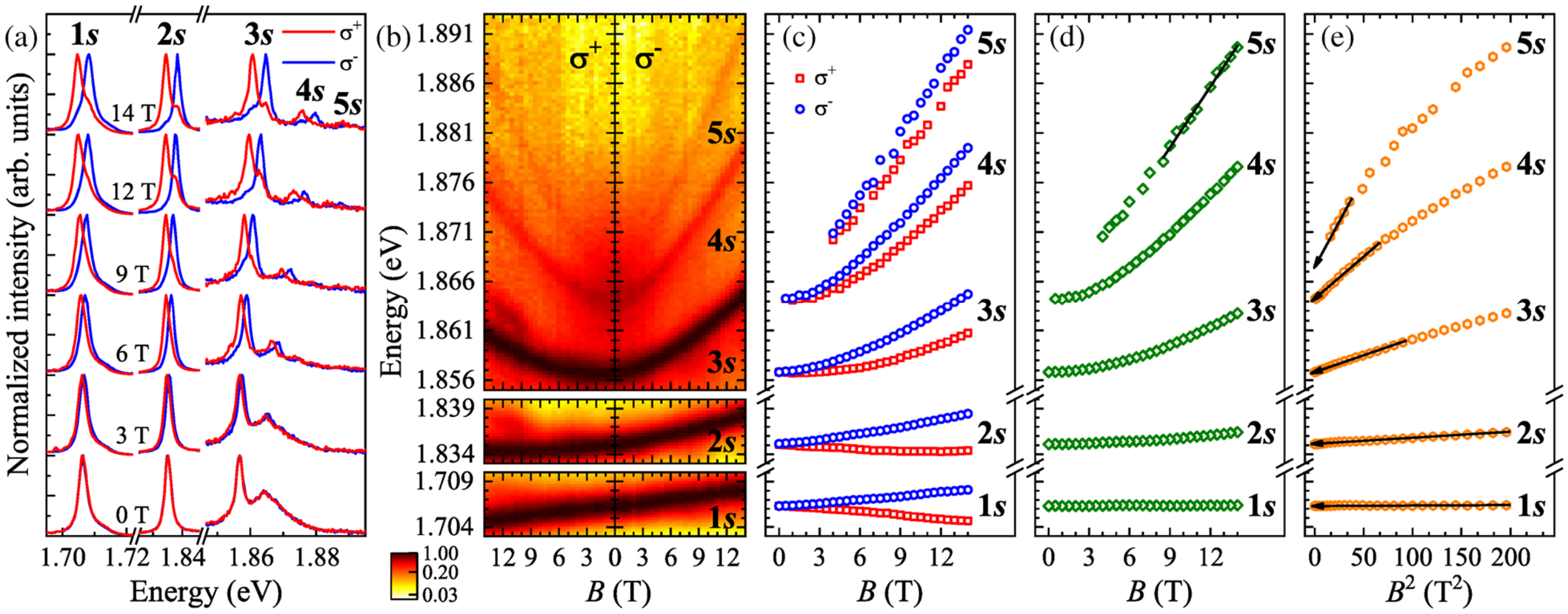

**Figure 18.** (a) and (b) Circular-polarization-resolved magneto-PL spectra of an hBN-encapsulated $WSe_2$ monolayer under out-of-plane magnetic fields of up to 14 T. (c) Magnetic-field-dependent energies of the exciton states up to $N = 5$ for the two polarizations showing valley Zeeman splitting of the Rydberg states. (d) and (e) Mean exciton energies versus $B$ and $B^2$ respectively. 1s and 2s states show quadratic diamagnetic shifts throughout the field range while 3s to 5s follow a quadratic behavior in low magnetic fields. This represents the low-field limit. 5s state shows a linear blue shift above about 9 T suggesting a crossover to the high-field limit. Reproduced with permission from Molas et al., Phys. Rev. Lett. 123, 136801 (2019). Copyright 2019 American Physical Society.

However, for 2D 2$s$, 3$s$ and 4$s$ excitons, the diamagnetic shifts are expected to be larger by a factor of 39, 275, and 1029 respectively, compared to the 2D 1$s$ exciton[50]. The measured diamagnetic shifts are used to determine the exciton-reduced mass, binding energy, and/or radius. The reduced mass determined this way can be compared with the high-field limit case described in the previous section.

For a 2D exciton in a GaAs quantum well, $\Delta E^{2D}_{dia}(1s)$ is about 2 meV at $B = 10$ T, and 7 meV at $B = 20$ T[45]. However, in a layered semiconductor such as a monolayer $WS_2$, this is extremely small for the A excitons (< 1.5 meV at 65 T)[142]. This is because the exciton reduced masses in TMDCs are much larger than in III-V semiconductors (0.15 in monolayer $WS_2$[221] compared to 0.04 – 0.08 for heavy hole excitons in GaAs QWs[46,47]), while their out-of-plane dielectric constants $\kappa$ are smaller (4.13 in monolayer $WS_2$[289] versus 12.5 in GaAs[46]). Therefore, one needs very large magnetic fields (still staying within the low-field limit) to measure the diamagnetic shifts of ground-state excitons in TMDCs with a reasonable accuracy[142]. For this reason, the diamagnetic shifts were not measured in monolayer TMDCs until 2016[142]. Examples of the measured $B^2$-dependent diamagnetic shifts in hBN-encapsulated $WSe_2$ are shown in Figures 17(b)[129] and 18(d,e)[130]. For bulk TMDCs, $\kappa$ is large (e.g. 14.5 for bulk $MoS_2$, and expected to be of the same order for other TMDCs[289]) resulting in larger diamagnetic shifts[142,143,290]. The exciton-reduced masses determined for A excitons in TMDCs following a combined analysis of the exciton energies in the low-field and high-field limits are compiled from various works in Table 1. The RMS radii of the excitons are noted in Table 2. A change of the dielectric environment around the monolayer affects the exciton binding energy and RMS radius, resulting in a different diamagnetic shift[286].

**Table 2.** RMS radii of the ground and excited states of A and B excitons ($X_A$ and $X_B$) in various TMDC crystals determined using magnetooptics by measuring diamagnetic shifts in the low-field limit ($\gamma \ll 1$). The references are mentioned along with the numbers as well. In addition to these, Wang et al. have recently measured the RMS radii of $Ns$ Rydberg states up to $N = 11$ where they get the RMS radius of 11$s$ state equal to 214 nm[131].

| Material | $X_A$ (nm) | | | $X_B$ (nm) |
|---|---|---|---|---|
| | 1s | 2s | 3s | 1s |
| 1L $WS_2$ on $SiO_2$ | 1.53[142] | - | - | 1.16[142] |
| 1L hBN/$MoS_2$/hBN | 1.2[202] | - | - | - |
| 1L hBN/$MoSe_2$/hBN | 1.1[202] | - | - | - |
| 1L hBN/$WS_2$/hBN | 1.8[202], 2[287] | 5 – 8[287] | - | - |
| 1L hBN/$WSe_2$/hBN | 1.7[202], 1.7[129], 1.75[131], 2.7[291] | 6.6[129], 6.8[131], 7.7[291] | 14.3[129], 15.45[131] | - |

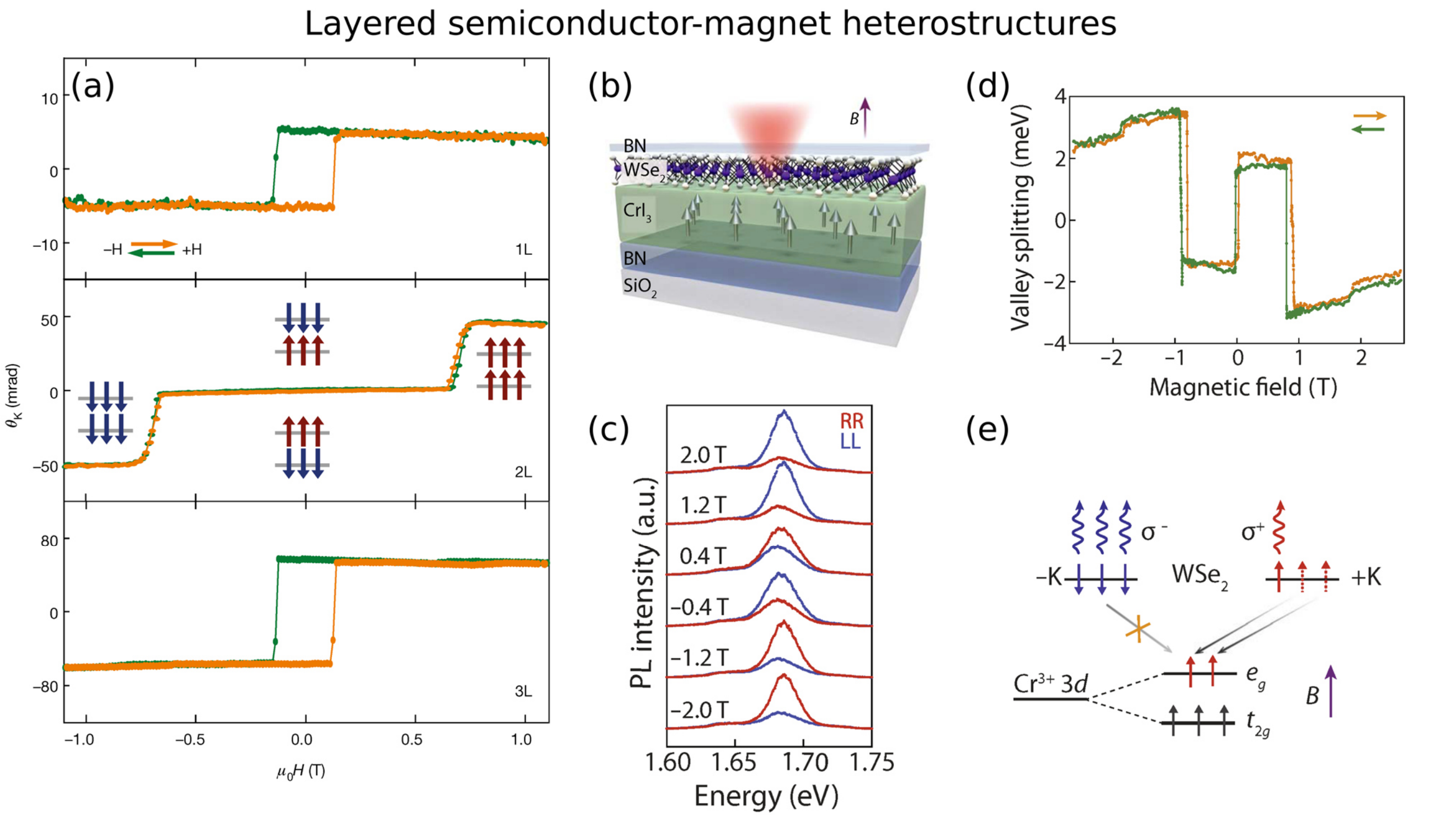

**Figure 19.** (a) Hysteresis loops of monolayer (1L), 2L and 3L thick layered magnet $CrI_3$, measured using MOKE spectroscopy. $\theta_K$ is the measured Kerr rotation. (b) Heterostructure made from 10 nm $CrI_3$ and monolayer $WSe_2$ sandwiched between thin layers of hBN. An exchange field of about 13 T has been detected at the heterojunction. (c) Magneto PL spectra of trions in monolayer $WSe_2$ in (b) in the two excitation/detection configurations RR and LL, where R and L denote right and left circularly polarized light. Strong valley polarization is observed under relatively small magnetic fields, arising from the exchange interaction at the heterojunction. (d) Valley Zeeman splitting between $K^{\pm}$ valleys of the monolayer $WSe_2$ derived from the magneto PL measurements. (e) Spin-orientation-dependent charge hopping between monolayer $WSe_2$ and thin $CrI_3$, resulting in polarized PL response as a function of magnetic field. (a) is reproduced with permission from Huang et al., Nature **546**, 270 (2017) (copyright 2017 Springer Nature), (b) – (e) are reproduced with permission from Zhong et al., Sci. Adv. **3**, e1603113 (2017) (copyright 2017 American Association for the Advancement of Science)

| | | | | |
|---|---|---|---|---|
| 1L hBN/$MoTe_2$/hBN | 1.3[202] | - | - | - |
| Bulk $WSe_2$ on $Al_2O_3$ | 2.0[144] | 4.3[144] | | |
| Bulk $MoTe_2$ on $SiO_2$ | 1.2[143] | 4.8[143] | - | - |

The experimentally obtained reduced masses are a matter of debate since discrepancies are found compared to *ab initio* calculations[130,163,202,221]. The disagreement is the strongest for $MoSe_2$ (up to 70% deviation) where the calculations predict a value of $0.27\,m_0$[163]. For other materials, the disagreement is of the order of $10-15\,\%$ which might be considered reasonable.

**2.5 Binding-energy enhancement.** The binding energy of the exciton (ground and excited states) also increases with magnetic field as $\sqrt{B}$ due to an enhanced overlap between electron and hole states under a magnetic field[38,51,272]. However, this effect is normally not noticed in the data since the other effects such as the $B^2$-dependent diamagnetic shift and $B$-dependent Landau quantization are more dominant. Therefore, it has been neglected in all works on TMDCs known to the author so far. It might be useful in the near future, to quantify and understand the extent of this effect theoretically for TMDCs so as to understand if it could be neglected. This could be important to check if it contributes to the discrepancies between the experimentally deduced exciton reduced masses and the theoretical calculations[130,163,202,221].

## 3. LAYERED MAGNETIC MATERIALS AND OUTLOOK

Research in the area of layered materials got another boost in 2017 with the discovery of a ferromagnetic ordered state in monolayers of Cr-trihalide $CrI_3$ (Curie Temperature $T_c$ = 45 K)[113], and bilayers of $Cr_2Ge_2Te_6$ ($T_c$ = 30 K)[115]. Here, polarization modulation and Sagnac interferometry based MOKE spectroscopy was used, respectively, for detecting ferromagnetic ordering as a function of temperature[113]. Since then, many other 2D ferromagnets such as other chromium trihalides i.e. $CrCl_3$ and $CrBr_3$[109], and more e.g. $Cr_2Ge_2Te_6$[115], $VSe_2$[116], and $Fe_3GeTe_2$[117] have been discovered. While chromium trihalides are semiconductors with *optical* band gaps in the visible electromagnetic spectral region[110,292], $VSe_2$ and $Fe_3GeTe_2$ are metallic[107,116,117]. Another emerging family of 2D magnetic semiconductors are antiferromagnets of the type $MPX_3$ ($M$ = Mn, Fe, V, Zn, Co, Ni, Cd, Mg; $X$ = S, Se)[111,112]. Many surprises are unraveled when a layer-thickness-dependent ordering is investigated. For instance, in the case of $CrI_3$, a monolayer and a trilayer are ferromagnetic, but a bilayer shows antiferromagnetic ordering (see Figure 19(a))[109]. The ferromagnetic order is restored in bulk $CrI_3$. $CrBr_3$ exhibits ferromagnetism irrespective of its layer thickness[109]. Recently, magneto-Raman spectroscopy has been used to identify magnons in the layered antiferromagnets $FePS_3$[293], and $MnPS_3$[294]. This approach could be extended to other layered magnetic materials to study magnetic excitations. For a summary of developments in this area of the research, the reader is referred to other recent reviews[107–110,295,296].

It was also shown in 2017 that an exchange interaction between $CrI_3$ thin film and a $WSe_2$ monolayer in a heterostructure could be used to control the valley Zeeman splitting and valley polarization in $WSe_2$ using small external magnetic fields (see Figure 19 (b-e))[114]. The authors reported an exchange-induced equivalent out-of-plane magnetic field of 13 T at the heterojunction. This provides unprecedented opportunities for exploring spin-valley physics of quasiparticles in TMDCs and device applications under external magnetic fields of practical importance (<1 T). Many future directions could be envisioned. For instance, the intervalley phase could be controlled by substantial amounts under small magnetic fields in such a heterostructure. Previously, such experiments required large magnetic fields up to 25 T[96,97]. Room-temperature control of intervalley phase and population under low-magnetic fields would be an essential requirement for a valleytronic gate. As another example, a suitable heterostructure between a 2D magnet and a TMDC could be used to create an in-plane exchange field and brighten the dark excitons using small external fields. Dark-exciton brightening has been proposed to be useful in chemical sensing applications[297]. Uniaxial strain on the layered magnets could be used to tweak the magnetic properties such as changing the orientation of the easy axis or the Curie/Néel temperature for future 'straintronic' devices[298,299]. Overall, the author believes that the magnetooptical methods would play an important role in the 2D-materials-based future technologies.

For unlocking the full potential of magnetooptics in layered materials, it would be essential to perform modulation magnetospectroscopy (e.g. magnetocircular dichroism, MOKE, Faraday rotation etc.). Although monochromatic MOKE has been used for studying magnetism in 2D materials, the use of modulation techniques to detect valley-Zeeman splitting effects on quasiparticle resonances such as excitons is scarce[152]. These methods are capable of measuring very low Zeeman splittings (few μeV)[150,152,300,301]. Furthermore, they provide a non-destructive way to study magnetic ordering in 2D magnetic materials. However, one requires to perform wavelength-dependent measurements for comparison with *ab initio* theory[110]. The bottleneck could be that it takes a long time (sometimes many hours) for measuring one spectrum. Keeping the micron size sample spots stable over such long measurement times could be challenging in free-space optics setups under magnetic fields. It is noteworthy that fiber-based setups which are normally used for magnetoreflectance/transmittance or PL measurements are not very suitable for modulation spectroscopy[87,104,148]. In future works, these methods are expected to pave a way to perform Zeeman spectroscopy on quasiparticle resonances in TMDCs under tiny magnetic fields. It would be important to check if the magnetic-field-dependent linearity of the valley Zeeman splittings of quasiparticles in TMDCs persist to very low fields ($< 2$ T) where no reliable data is available so far. Furthermore, modulation Zeeman spectroscopy under relatively large magnetic fields (10 T and more) could be vital to shine light on the physics of highly-excited Rydberg states of excitons whose signatures in non-modulated spectra remain feeble such as in Figure 17(b and c).

In summary, this perspective highlights the progress and future prospects of many scientific questions of fundamental importance in the area of 2D semiconductors using magnetooptical spectroscopy. Although the area of research is relatively recent (5-6 years), a rapid progress has been witnessed. Many puzzles, sometimes decades old, have been solved, and there is an enormous scope for future research from both theoretical and experimental perspectives. Overall, the author firmly believes that non-destructive magnetooptical techniques have a strong potential to be useful in future fundamental explorations of 2D semiconductors and magnetic materials, as well as device applications in many emerging fields such as spintronics, valleytronics, straintronics, twistronics, and quantum information and computation.

**Acknowledgements.** The author gratefully acknowledges countless discussions on magnetooptics of semiconductors with Marek Potemski over the course of many years and has benefitted immensely from discussions with Rudolf Bratschitsch, Xavier Marie, Thorsten Deilmann, Alexey Chernikov, Tobias Korn, Clément Faugeras, Sandip Ghosh, Andreas Stier, Maciej Molas, and Cedric Robert while writing this paper. The work was supported by the DFG projects AR 1128/1-1 and AR 1128/1-2. Help from Shreya Agrawal during manuscript preparation is gratefully acknowledged.

**Data availability statement.** The data that support the findings of this study are available from the corresponding authors of the original manuscripts discussed.

**References.**